\documentclass[journal=jctcce,manuscript=article]{achemso}

\usepackage{chemformula} 
\usepackage[T1]{fontenc} 

\usepackage{graphicx}
\usepackage{amsmath}
\usepackage{amssymb}
\usepackage{braket}
\usepackage{multirow}
\usepackage{soul}
\usepackage{siunitx}

\newcommand{\vect}[1]{\ensuremath{\boldsymbol{#1}}}

\newcommand{\pd}[2]{\frac{\partial #1}{\partial #2}}

\author{Maarten Cools-Ceuppens}
\affiliation{Center for Molecular Modeling (CMM), Ghent University - Technologiepark-Zwijnaarde 46, B-9052 Gent, Belgium}

\author{Joni Dambre}
\affiliation{IDLab, Electronics and Information Systems Department, Ghent University - imec, Technologiepark-Zwijnaarde 126, B-9052 Gent, Belgium}

\author{Toon Verstraelen}
\email{toon.verstraelen@ugent.be}
\affiliation{Center for Molecular Modeling (CMM), Ghent University - Technologiepark-Zwijnaarde 46, B-9052 Gent, Belgium}

\title{Modeling electronic response properties with an explicit-electron machine learning potential}

\begin{document}

	\begin{tocentry}
	\includegraphics{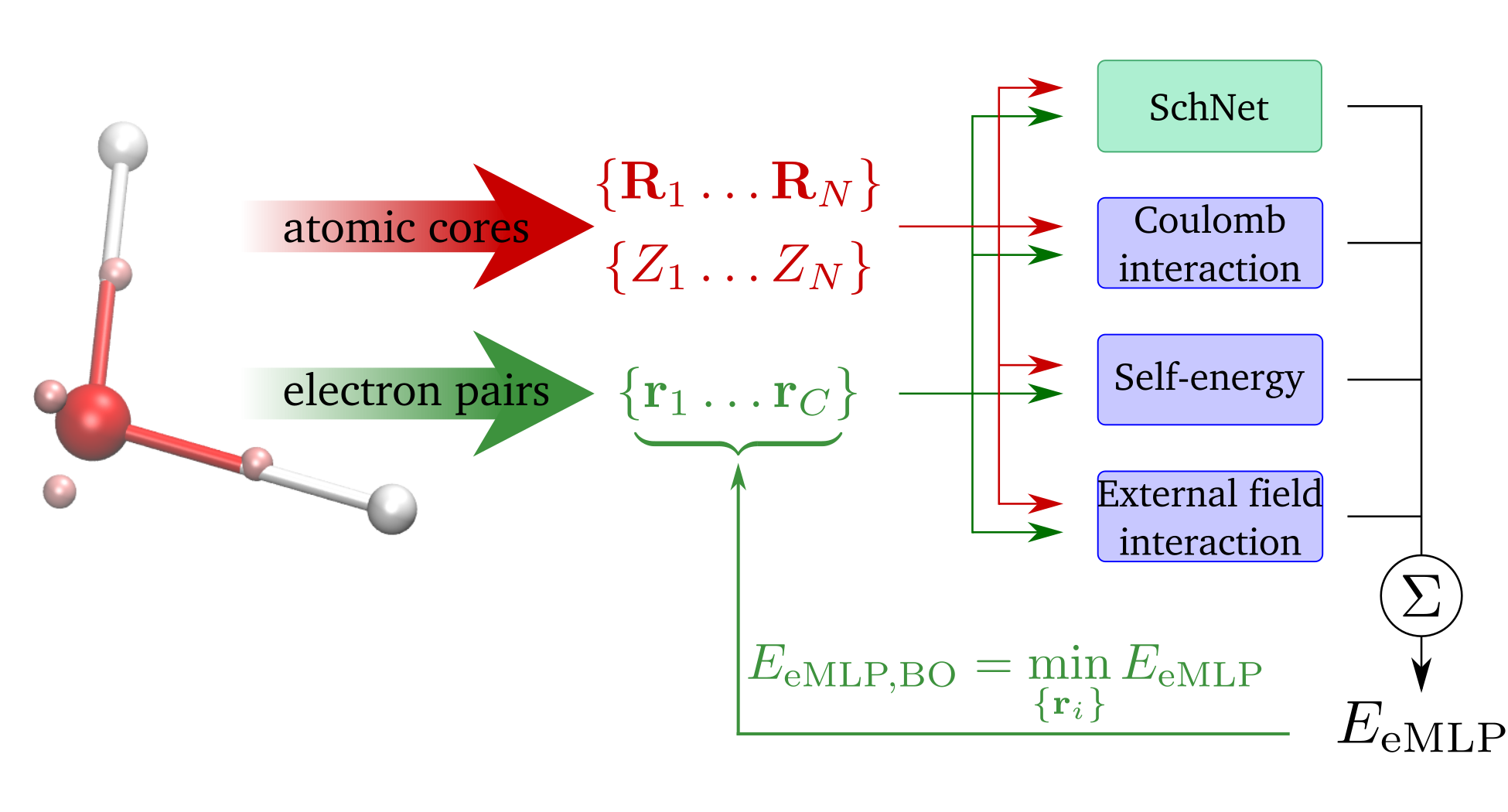}
	\end{tocentry}

	\begin{abstract}
	    Explicit-electron force fields introduce electrons or electron pairs as semi-classical particles in force fields or empirical potentials, which are suitable for molecular dynamics simulations. Even though semi-classical electrons are a drastic simplification compared to a quantum-mechanical electronic wavefunction, they still retain a relatively detailed electronic model compared to conventional polarizable and reactive force fields. The ability of explicit-electron models to describe chemical reactions and electronic response properties has already been demonstrated, yet the description of short-range interactions for a broad range of chemical systems remains challenging. In this work, we present the electron machine learning potential (eMLP), a new explicit electron force field where the short-range interactions are modeled with machine learning. The electron pair particles will be located at well-defined positions, derived from localized molecular orbitals or Wannier centers, naturally imposing the correct dielectric and piezoelectric behavior of the system. The eMLP is benchmarked on two newly constructed datasets: eQM7, a extension of the QM7 dataset for small molecules, and a dataset for the crystalline $\beta$-glycine. It is shown that the eMLP can predict dipole moments, polarizabilities and IR-spectra of unseen molecules with high precision. Furthermore, a variety of response properties, e.g.\ stiffness or piezoelectric constants, can be accurately reproduced.
	\end{abstract}

	\section{Introduction}

	A central problem in computational chemistry is finding approximate, yet sufficiently accurate, solutions for the quantum-mechanical electronic structure problem, given a configuration of nuclei in a molecule or a condensed system. Many microscopic properties of such a system can be derived once the wavefunction is solved, such as the potential energy surface, a molecular dipole moment, etc. Usually, the calculation of the electronic wavefunction is merely a required intermediate step towards those properties of interest. For this reason, many force field models have been developed, which compute properties of interest directly, bypassing the electronic wavefunction. \cite{warshel_polarizable_2007,harrison_review_2018} Their main advantage is a drastic reduction in computational cost, bringing much larger atomistic systems and their dynamics at longer time scales in reach of computer simulations. Still, (approximate) electronic structure calculations are widely used despite their higher computational cost. In general, they predict properties more accurately for a broad range of systems and they rely less on empirically adjusted model parameters. In addition, the electronic wavefuction and its response to external stimuli gives access to many properties of interest. In force fields, such electronic properties are not trivially available, due to the absence of a detailed model of electronic structure.

	Despite the lack of an electronic wavefunction, force-field models can incorporate electronic features to some degree, such that a subset of the electronic properties can be derived. The most common feature is a fixed partial charge for each atom, which is mainly used to describe electrostatic properties and the corresponding long-range interactions. \cite{dykstra_electrostatic_1993,neves-petersen_protein_2003} Atomic partial charges are essentially a coarse-grained description of a frozen electron density. Polarizable force fields go beyond static charge distributions by also modeling the change in electron density due to an applied external field. \cite{Cieplak_2009, doi:10.1146/annurev-biophys-070317-033349} This is typically accomplished by introducing variables in the atomic multipole expansions, e.g. induced dipoles, which are solved by an energy minimization. Most polarizable force fields rely on a linear-response approximation, optionally with non-linear corrections, \cite{rappe_charge_1991} limiting their applicability to small electronic rearrangements. A less appreciated limitation of most polarizable force fields is that they only attempt to approximate changes in electron density. This is problematic because a change in electron density alone cannot describe the change in macroscopic polarization of a periodic system. \cite{resta_theory_2007} Simply put, under periodic boundary conditions, a change in electron density is insufficient to derive from where to where electrons have moved, hampering a sound definition of macroscopic polarization. Unless additional assumptions are made, polarizable force fields inherent this ambiguity. Fortunately, most polarizable models avoid this difficulty by clearly specifying from where to where charge is displaced. For example, in a Drude oscillator \cite{doi:10.1021/acs.chemrev.5b00505} or induced dipole \cite{thole_dipole_1981,doi:10.1021/ja00181a017}, model polarization is local within each atom. However, for models with fluctuating atomic charges, e.g. EEM, \cite{mortier_eem_1986} QEq \cite{rappe_charge_1991,doi:10.1002/jcc.10355} or machine learned models for partial charges, \cite{doi:10.1021/acs.jcim.0c01071,Ko2021} the polarization of periodic models remains ambiguous, which will be shown in the next paragraph. A completely different approach to incorporate electronic degrees of freedom in force fields, are the so-called semi-classical or explicit electron force fields. \cite{C7SC01181D} These models introduce local electrons or electron pairs as negatively charged particles, in addition to the nuclei or positively charged atomic cores. Because all particles have a fixed charge, just as in quantum-mechanical electronic structure methods, they have no issues defining the polarization of a periodic model. Furthermore, by allowing electron (pair) particles to migrate away from their original bound configuration, processes such as ionization, redox reactions and charge transport, can be in principle described. \cite{doi:10.1063/1.3688228, Kale2012, https://doi.org/10.1002/jcc.23612, doi:10.1098/rspa.2015.0370, doi:10.1021/acs.jpcb.6b02576} Because these processes depart manifestly from a linear response regime, they remain challenging for conventional polarizable force fields. This work is motivated by the appealing prospects of explicit electron models and explores new approaches to develop such models.

\begin{figure}
		\begin{center}
			\includegraphics[width=\textwidth]{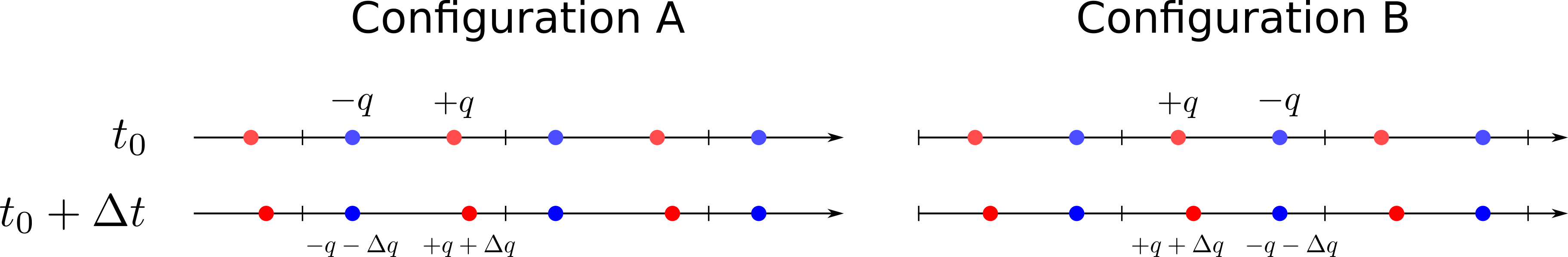}
			\caption{A one-dimensional system with lattice length $a$, containing two charged particles. At $t_0$, the particles have a charge of $-q$ and $+q$ and are separated by $a/2$. At $t_0 + \Delta t$, the positive particle has moved a distance $\Delta r$ and gained a charge $\Delta q$ while the negative particle stays fixed but loses $\Delta q$ in charge. Both configurations describe the same system but the origin of unit cell is translated over a distance $a/2$.} \label{fig:charge_transfer}
		\end{center}
	\end{figure}

	Models with geometry-dependent atomic charges have been developed for several decades and are widely employed for molecular simulations. \cite{mortier_eem_1986, rappe_charge_1991, doi:10.1002/jcc.10355, nistor_sqe_2006, warren_origin_2008, nistor_dielectric_2009, PhysRevB.92.045131, PhysRevB.95.104105, Ko2021, doi:10.1021/acs.jctc.0c00217, doi:10.1021/acs.jcim.0c01071} To the best of our knowledge, it was never reported previously that such models are inconsistent with the modern theory of polarization and are therefore problematic for modeling dielectric systems. We will clarify this issue with a simple one-dimensional example. Note that the issue shown here is more general than the metallic polarizability scaling of charge equilibration models, \cite{warren_origin_2008, nistor_dielectric_2009} because we do not assume partial charges are found by charge equilibration. It is well-known in the modern theory of polarization that the dipole vector itself is not well defined for a periodic system and that only the change in polarization has a physical meaning and can be measured experimentally. This was illustrated by Spaldin \cite{SPALDIN20122} for a one-dimensional lattice with fixed charges. Here, we take the same example and allow charges to fluctuate when they are displaced, to demonstrate that this results in an ill-defined change in polarization. Consider a one-dimensional lattice with lattice length $a$, as depicted in Figure \ref{fig:charge_transfer}. Two particles are present with opposite charges $+q$ and $-q$, separated by a distance $a/2$ at the initial time $t_0$. At a later time $t_0 + \Delta t$, the positively charged particles moves a distance $\Delta r$ to the right while gaining a additional charge $\Delta q$. In configuration A, the unit cell is chosen such that the negative particle starts at $a/4$ and the positive particle at $3a/4$. Exactly the opposite is true in configuration B but they both describe the same system when periodic boundary conditions are taken into account. In configuration A, the change in dipole per lattice length is
	\begin{align}
	    \Delta p &= \frac{1}{a} \sum_i q_i(t_0 + \Delta t) r_i (t_0 + \Delta t) - \frac{1}{a} \sum_i q_i(t_0) r_i (t_0) \nonumber \\
	    &= (q + \Delta q) \left( \frac{3}{4} + \frac{\Delta r}{a} \right) + (-q - \Delta q) \frac{1}{4} - q \frac{3}{4} + q \frac{1}{4} \nonumber \\
	    &= (q + \Delta q) \frac{\Delta r}{a} + \frac{1}{2} \Delta q
	\end{align}
    while for configuration B, the same equation yields
	\begin{align}
	    \Delta p &= (q + \Delta q) \left( \frac{1}{4} + \frac{\Delta r}{a} \right) + (-q - \Delta q) \frac{3}{4} - q \frac{1}{4} + q \frac{3}{4} \nonumber \\
	    &= (q + \Delta q) \frac{\Delta r}{a} - \frac{1}{2} \Delta q
	\end{align}
	Hence, the change in polarization depends on the chosen definition of the unit cell, which is a major concern since this quantity can be measured experimentally. Note that the problematic term $\pm \Delta q /2$ is related to the charge transfer. Within the unit cell of configuration A, a charge of $\Delta q$ is transferred to the right over half the lattice length, while in configuration B, the charge moves to the left. The change in polarization only becomes well-defined if there are no fluctuating charges $\Delta q = 0$ or when one unambiguously defines from where to where charge is transferred, such as in the split-charge equilibration model. \cite{nistor_sqe_2006}

	A variety of explicit electron force fields have been developed. The electron force field\cite{PhysRevLett.99.185003, doi:10.1063/1.3272671} (eFF) and its successor eFF-ECP\cite{XIAO2015243, PhysRevLett.108.045501} (eFF with effective core potentials) are mainly established to model materials under extreme conditions, whether it are high pressures or high temperatures. The LEWIS\cite{doi:10.1063/1.3688228, Kale2012} force field is capable of simulating liquid water and its dielectric response. It has been extended to LEWIS$\bullet$ \cite{https://doi.org/10.1002/jcc.23612, doi:10.1098/rspa.2015.0370, doi:10.1021/acs.jpcb.6b02576} to incorporate 2$p$ and 3$p$ elements and diatomic molecules and shows promising results in predicting electron affinities and ionization potentials. More recently, explicit electron extensions to the reactive force field ReaxFF have been developed. The inclusion of electrons or holes as additional particles is realized in eReaxFF\cite{doi:10.1021/acs.jctc.6b00432}, whereas in ReaxFF/C-GeM \cite{doi:10.1021/acs.jpclett.0c02516, doi:10.1021/acs.jpclett.9b02771} (ReaxFF and the coarse-grained electron model), each atom is characterized by a positive core and negative shell (not necessarily at the same location) and modeled as interacting Gaussian charges. All these methods share the same viewpoint of the electron as a pseudo-classical particle but differ in a variety of aspects. One can model all electrons, both the core and valence electrons, or only model the valence electrons. In the latter scenario, the core electrons are simply frozen at the positions of the nuclei and not treated explicitly. Next, one can choose to separate the spin up and spin down electrons as different particles or group a spin-up and spin-down electron as an electron pair, represented by a single particle with charge -2\,e, which reduces the amount of particles, lowers the model complexity and improves computational efficiency. Also the parameters in explicit electron force fields have been estimated in several different ways. Unless the models are trained directly to terms involving the positions of the electrons, like the dipole vector, the electrons will be located at positions which are a consequence of the fit and do not have a direct relation to the electronic wavefunction. One of the innovations in this work is that the electron pair particles will be positioned at centers of localized molecular orbitals. \cite{PhysRev.140.A1133} These centers have a reasonable similarity to Lewis structures and provides ample microscopic data for training. Furthermore, by placing -1\,e charges at centers of localized orbitals, one reproduces the molecular dipole moment exactly, guaranteeing proper long-range electrostatics.

	A major difficulty in the process of developing explicit electron force fields, is the characterization of the short-range interactions. At long distances, the classical Coulomb electrostatics are an adequate approximation for the interaction between charged particles but a short distances quantum effects may dominate. Electrons are fermions for which the Pauli exclusion principle is valid, resulting in exchange interactions, where its effect on explicit electron force fields have been already investigated in detail\cite{C6CP06100A}. In this work however, we do not attempt to derive the short-range interactions from theoretical derivations or via heuristic approximations. Instead they will be modeled with machine learning to avoid any assumptions on their functional form. In general, machine learning force fields\cite{doi:10.1021/acs.chemrev.0c01111, huang2020ab, doi:10.1021/acs.chemrev.0c00868} try to learn the relation between the geometry and chemical species in the system and the total energy and its derivatives. No physical insight is required, only a vast amount of ab-initio data is necessary to fit the potential energy surface (PES). Most of the machine learning models can be subdivided in a few classes: the neural networks\cite{PhysRevLett.98.146401, doi:10.1063/1.3553717, C6SC05720A} or the message-passing\cite{gilmer2017neural} (deep) neural networks (MPNN)\cite{doi:10.1063/1.5019779, Zubatyukeaav6490, doi:10.1063/1.5011181, doi:10.1021/acs.jctc.9b00181, klicpera2020fast, schutt2021equivariant} and kernel-based methods\cite{Chmiela2018, PhysRevLett.104.136403, PhysRevB.87.184115, doi:10.1063/1.5126701, doi:10.1063/1.5053562}. In essence, a local representation of every atom in a multidimensional vector space is defined as a feature (for the kernel-based methods) or learned (for the message passing neural networks). This representation encodes the chemical environment around every atom in a certain cutoff radius and serves as the input to predict the atomic energies. We will utilize the SchNet\cite{doi:10.1063/1.5019779, doi:10.1021/acs.jctc.8b00908} deep neural network to model the short-range interactions in an explicit electron force field. It is a well-proven and benchmarked\cite{doi:10.1063/5.0038516, doi:10.1021/acs.jctc.0c00347} architecture which performs equally well on relevant benchmarks compared to other state-of-the-art machine learning force fields, given that enough data is available. Furthermore, the prohibitive scaling of the number of features per atom, with respect to total amount of chemical elements in the system, is avoided because a MPNNs learn their own representation.

	Our new model is not only an refinement upon existing explicit electron force fields but is also an innovation in the treatment of electrostatic and polarization interactions in machine learning force fields. Typically, long-range interactions are modeled by learning partial charges for every atom based on the local representation, after which the charges then interact with classical electrostatics. These are non-polarizable machine learning force fields because the charges only depend on the local environment and are insensitive to electric fields from more distant charge distributions. This is addressed in so-called fourth generation neural network potentials\cite{doi:10.1021/acs.chemrev.0c00868}. Another concern is that charges directly predicted by neural networks have an incorrect total charge and must be shifted \textit{ad hoc}.\cite{doi:10.1021/acs.jctc.9b00181} In the context of neural networks, only recently some advancements have been made to address these charge-issues. Models based on the charge equilibration neural network technique (CENT)\cite{PhysRevB.92.045131, PhysRevB.95.104105, Ko2021} do predict atomic electronegativities, which are employed to optimize the charges by minimizing the electrostatic energy. In the BpopNN model\cite{doi:10.1021/acs.jctc.0c00217}, electronic populations are introduced, which serve as extra input variables in the neural network, where the optimal values are again calculated by minimizing the energy. In AIMNet-ME\cite{Zubatyuk2020}, a total charge of -1\,e, 0\,e or 1\,e can be imposed as an extra feature in an MPNN, while the electron passing neural network (EPNN)\cite{doi:10.1021/acs.jcim.0c01071} iteratively updates partial charges in its message passing network, while constraining the total charge. Total charge and spin are constrained and serve as extra input in SpookyNet\cite{unke2021spookynet}, which also includes non-local interactions by using self-attention and analytic long-range corrections. The influence of electric fields are taken into account in FieldSchNet\cite{gastegger2020machine} by learning vectorial representations per atom and coupling it with the external field by taking scalar products. For kernel-based methods, progress has been made by incorporating the long-distance equivariant (LODE)\cite{D0SC04934D} representation to describe long-range effects. The majority of these efforts to incorporate electronic polarization in machine-learned potentials rely on environment-dependent fractional charges. As shown above, this leads to non-trivial difficulties when describing the polarization of periodic systems. This motivated us to explore explicit electrons as an alternative approximate representation of the electronic structure.

	In this work, we will present a new explicit electron force field, which we will call the electron machine learning potential (eMLP), where the short-range interactions are learned via machine learning. For now, only electron pairs will be considered by grouping up the spin-up and spin-down electrons in a single particle. Furthermore, only valence electron pairs are considered as a starting point. This simplifies the methodology and training of the neural network. In addition, this also improves the computational efficiency when making predictions. The electron pairs will be located at well-defined positions, derived from localized molecular orbitals. This will naturally impose the correct dipole moments, polarization or piezoelectric behavior of the system. We will study the ability of the model to predict polarizabilities and IR spectra of unseen small molecules. For periodic systems, we will focus on $\beta$-glycine as a case study, since piezoelectricity in biomolecules has gained a lot of attention in research\cite{Guerin2019} in recent years due to the possibly large piezoelectric strain responses. It is shown that the eMLP will make it possible to accurately reproduce stiffness, dielectric and piezoelectric constants for $\beta$-glycine. Finally, to enable stable molecular dynamics (MD) simulations, data-augmentation will be introduced. During the training phase, additional out-of-equilibrium electron pair positions will be generated since they are poorly sampled by conventional techniques.

	Sec. \ref{sec:methodology} discusses the mathematical structure of short- and long-range interactions in eMLP and the computational details localization procedure. Next, the databases will be introduced. Finally, we will explain how the model is trained, with or without data augmentation. In Sec. \ref{sec:results} the results will be discussed for eQM7, a dataset of small molecules, and crystalline $\beta$-glycine, after which the main conclusions and outlook will be presented in Sec. \ref{sec:conclusion}.



	\section{Methodology} \label{sec:methodology}

	In this section, the methodology followed in this work will be explained in detail. We will start by describing the particles in the eMLP, its overall structure and the energy decomposition. In the next two consecutive subsections, the long-range and short-range contributions will be defined. Subsequently, it is described how the electronic positions are extracted from ab initio calculations by making use of electron localization, both in periodic and non-periodic systems. Next, we introduce two new datasets, created for this work: eQM7, a dataset for the purpose of modeling a variety of small molecules and a dataset for $\beta$-glycine. In the section thereafter, data-augmentation will be discussed to overcome sampling deficiencies with respect to the electronic degrees of freedom. Finally, the cost function to train the eMLP will be presented.

	\subsection{Description of the model}

	\subsubsection{Overall structure}

	The eMLP treats the electrons as semi-classical particles in addition to the nuclei. In this work, only systems with an even number of electrons will be considered and all pairs of spin-up and -down electrons are grouped into a single pair particle with charge -2\,e, similarly to restricted closed-shell Hartree-Fock (RHF). The eMLP in this work is developed for systems comprising elements H, C, N and O.  We assume that the 1s core electron pair in C, N or O is located exactly on top of their corresponding nucleus, such that only the valence electron pairs need to be explicitly described. This will be validated in section \ref{sec:localization}. From this point onward, the term \textit{electron pair} will be reserved for valence electron pairs and the term \textit{atomic core} will be used to describe the ensemble of the nucleus and where applicable the core electrons. The terminology is summarized in Table \ref{tab:particle_types}.

	\begin{table}
		\begin{center}
			\begin{tabular}{c | c | c }
				 & atomic core & electron pair  \\ \hline
				symbol of position & $\vect{R}_i$ & $\vect{r}_i$ \\ \hline
				\multirow{2}{*}{charge [e]} & +1 for $Z_i = 1$ & \multirow{2}{*}{-2} \\
				& $Z_i - 2$ for $Z_i \in {6, 7, 8}$ & \\ \hline
			\end{tabular}
			\caption{An overview of the particles appearing in the eMLP and their symbols and charges. \label{tab:particle_types}}
		\end{center}
	\end{table}

\begin{figure}
		\begin{center}
			\includegraphics[width = 0.3\textwidth]{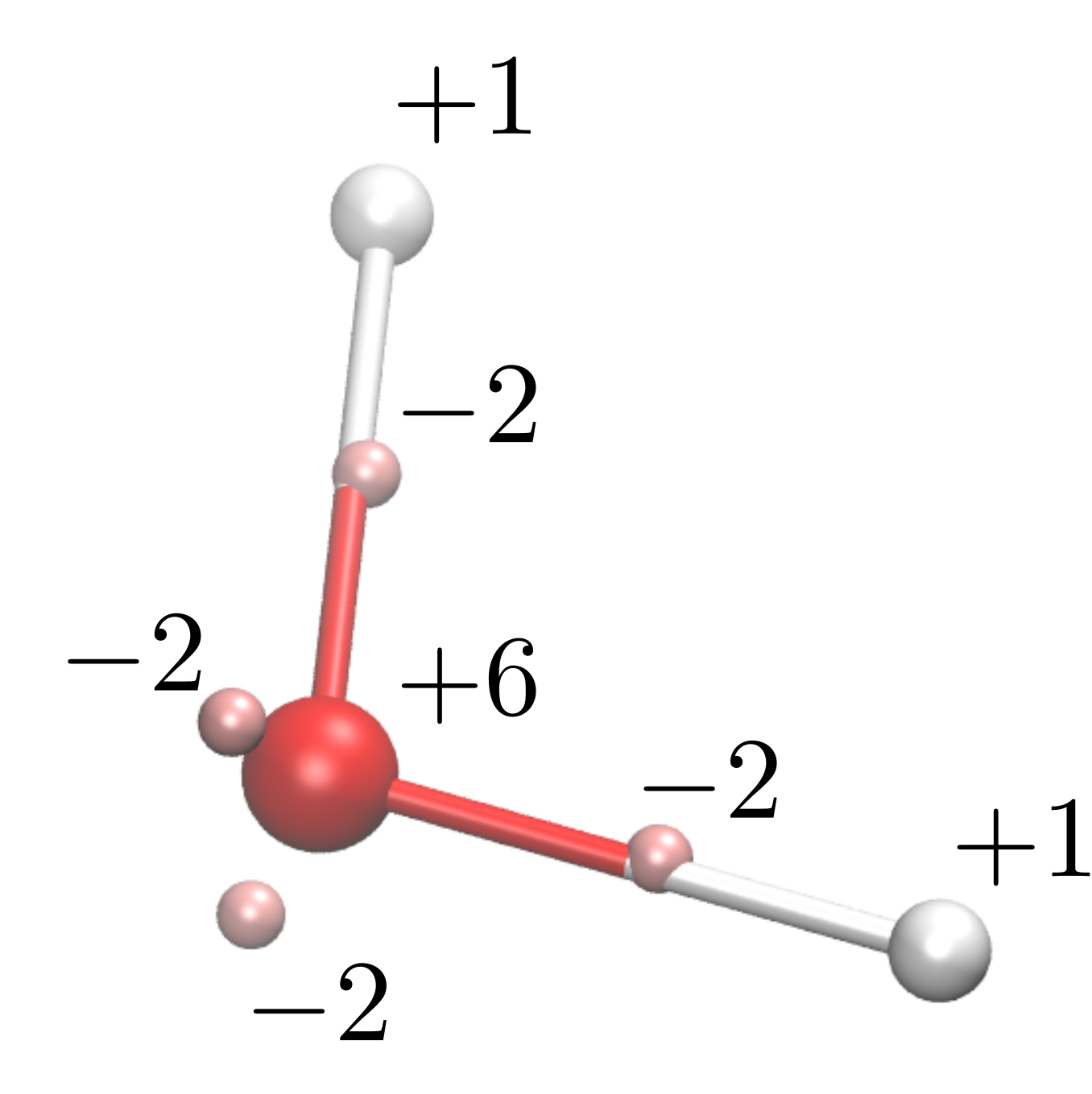}
			\caption{A single water molecule with 10 electrons, giving rise to four explicit electron pairs. The charge (in e) of each particle is indicated.} \label{fig:water_fb_charges}
		\end{center}
	\end{figure}

	As an example, consider a single water molecule (a 10-electron system) where the particles and their charges are visualized in Fig. \ref{fig:water_fb_charges}. Four electron pairs are explicitly described by the eMLP: two electron pairs participate in the bond between the oxygen and hydrogen atomic cores and another two give rise to the lone pairs of the oxygen atom. The final two electrons are core electrons which are not treated explicitly but are incorporated in the atomic core of oxygen, giving it a charge of $+6\,\mathrm{e}$.

	The eMLP models an extended potential energy surface (PES) including electronic degrees of freedom. Besides the positions of atomic cores $\vect{R}_i$ and their atomic numbers $Z_i$, electron pair positions $\vect{r}_i$ serve as inputs for the potential energy of the system: $E_\text{eMLP} = E_\text{eMLP}(\{\vect{R}_i\}, \{Z_i\} ; \{\vect{r}_i\})$.
	\begin{figure}
		\begin{center}
			\includegraphics[width = \textwidth]{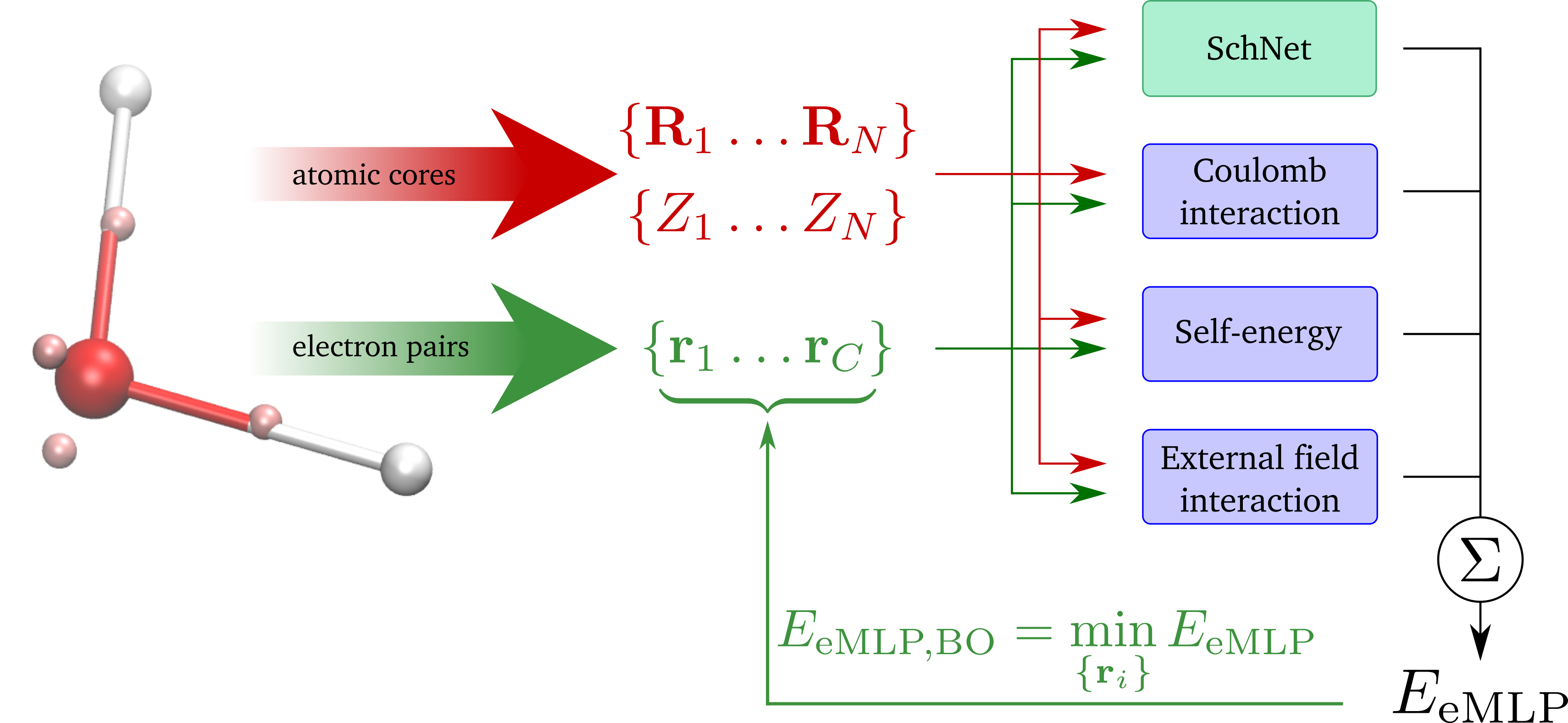}
			\caption{A schematic overview of the eMLP. Besides positions and species of the atomic cores, positions of the electron pairs are needed as inputs for the eMLP. The four building blocks of the eMLP are the short-range machine learning contribution (SchNet) and the three long-range contributions (the Coulomb interaction, self-energy and external field interaction). In MD simulations, the Born-Oppenheimer eMLP energy is needed, for which the energy is minimized with respect to the electron pair positions} \label{fig:model_architecture}
		\end{center}
	\end{figure}
	The extended PES is modeled by combining machine learning and classical long-range electrostatics.
    \begin{align}
        E_\text{eMLP} = E_\text{long-range} + E_\text{short-range}
    \end{align}

    The complex short-range interactions are modeled with SchNet\cite{doi:10.1063/1.5019779}, a deep neural network with over one million trainable parameters. The long-range interactions consist of the Coulomb repulsion or attraction between Gaussian charge densities centered at each particle, together with a self-energy term and the interaction with an external field. A schematic overview of the eMLP and its energy contributions is given in Fig. \ref{fig:model_architecture}. The energy partitioning has several advantages. First of all, machine learning force fields require no physical insight. This is helpful since the short-range interactions between the electron pairs and the atomic cores are non-trivial. Furthermore, making the long-range interactions parameter-free will help to overcome overfitting issues. The forces of the long-range interactions are after all harder to learn since they are generally smaller than those of the short-range contributions. Additionally, this removes the need for representative datapoints for the long-range interactions in the dataset as the total amount of possible chemical environments drastically increases at longer length scales. The partitioning ultimately increases the transferability of the eMLP. In the next two subsections, both interaction types are described in more detail.

    This extended PES already enables us to optimize geometries (both atomic cores and electron pairs) and to predict the dipole moments, polarizabilities, infrared (IR) spectra and more. However, molecular dynamics (MD) simulations require some extra considerations for the dynamics of the electron pairs. We follow a similar reasoning as in the Born-Oppenheimer approach: the electron pairs are significantly lighter than the atomic cores, such that in each step of the MD simulation the electronic positions are being relaxed,
	\begin{equation}
		E_\text{eMLP,BO}(\{\vect{R}_i\}, \{Z_i\}) = \min_{\{\vect{r}_i\}} E_\text{eMLP}(\{\vect{R}_i\}, \{Z_i\}, \{\vect{r}_i\}), \label{eq:minimization}
	\end{equation}
	resulting in a conventional PES, suitable for MD simulations. This closely resembles an SCF optimization for DFT calculations. The positions of the electron pairs for which the energy is minimized $\vect{r}_i^\text{eq}$ can be considered as a side-product in each MD step. In a system with $C$ electron pairs, the forces on the atomic cores are simply given by
	\begin{equation}
	\begin{aligned}
	    \vect{F}_{\text{BO},j}
	        &= -\pd{E_\text{eMLP,BO}(\{\vect{R}_i\}, \{Z_i\}}{\vect{R}_j} \\
	        &= -\pd{E_\text{eMLP}(\{\vect{R}_i\}, \{Z_i\}, \{\vect{r}_i^\text{eq}\})}{\vect{R}_j}
	    - \sum_{k=1}^{C} \pd{E_\text{eMLP}(\{\vect{R}_i\}, \{Z_i\}, \{\vect{r}_i^\text{eq}\})}{\vect{r}_k^\text{eq}}
	    \pd{\vect{r}_k^\text{eq}}{\vect{R}_j},
	\end{aligned}
	\end{equation}
	of which the second term is zero since the forces on the electron pairs,
	\begin{equation}
	    \vect{f}_j = -\pd{E_\text{eMLP}(\{\vect{R}_i\}, \{Z_i\}, \{\vect{r}_i^\text{eq}\})}{\vect{r}_j},
	\end{equation}
    are zero for the equilibrium positions.

	\subsubsection{Long-range interactions} \label{sec:long-range}

    The long-range energy in eMLP consists of three contributions:
    \begin{equation}
	    E_\text{long-range} = E_\text{Coulomb} + E_\text{self} + E_\text{ext}
    \end{equation}
    which will be explained in more details below.

	All the particles are modeled as Gaussian charge densities and they interact with each other through the Coulomb interaction. This results in the following energy contribution
	\begin{equation}
	    E_\text{Coulomb} = \frac{1}{2} \sum_{i \neq j} q_i q_j \frac{\text{erf}(\gamma ||\vect{r_i} - \vect{r_j}||)}{||\vect{r_i} - \vect{r_j}||}
	\end{equation}
	where the double sum runs over all particles, both atomic cores and electron pairs, having a charge of $q_i$ (see Table \ref{tab:particle_types} for the numerical value of charges). The parameter $\gamma$ appearing in the error function is inverse proportional to the width of the Gaussian charges and controls the distance for which the electrostatic potential is damped compared to point-charge interactions. In this work, the eMLP will use $\gamma = 0.3928$\,/\AA, which is chosen such that the energy between particles greater than the cutoff radius $r_\text{cutoff} = 4$\,\AA\ is approximately that of two point charges, while huge attractive or repulsive forces inside the cutoff radius are damped. Note that $\gamma$ is not intended to mimic the spatial extent of localized orbitals. It is solely designed to lower the magnitude of short-range electrostatic forces, as will be further explained below.

	The long-range energy also includes a self-energy term for each particle:
	\begin{align}
	    E_\text{self} & = \frac{\gamma}{\sqrt{\pi}}\sum_{i} q_i^2,
	\end{align}
	whose main purpose is to normalize the magnitude of the total electrostatic energy:
	\begin{equation}
	    E_\text{Coulomb} + E_\text{self} \approx 0 \label{eq:long-range_normalization}
	\end{equation}
	When charged particles are sufficiently far apart, $E_\text{Coulomb}$ would go to zero already. However, in all datasets in this work, charges are relatively close and $E_\text{Coulomb}$ is large in magnitude. In the limit of overlapping charge distributions, making use of $\lim_{x\rightarrow0}\operatorname{erf}(x)/x=2/\sqrt{\pi}$, the sum of the Coulomb and self energy becomes zero for neutral systems:
	\begin{equation}
	    E_\text{Coulomb} + E_\text{self} \approx \frac{\gamma}{\sqrt{\pi}} \sum_{i \neq j} q_i q_j + \frac{\gamma}{\sqrt{\pi}}\sum_{i} q_i^2 = \frac{\gamma}{\sqrt{\pi}}\left(\sum_i q_i\right)^2
	\end{equation}
	For all systems considered in this work, the sum of these two terms is much closer to zero than the Coulomb energy alone.

    By including the self-energy, not only the long-range forces are small in magnitude, but also the long-range energies. If they were not, the short-range interaction would need to cancel out these large forces and energies. For instance, a lone electron pair sitting around an oxygen atom at a distance of 0.3\,\AA\ will feel attractive forces up to 2000\,eV/\AA\ if their interaction was modeled via point charges. On the other hand, the average magnitude of the forces of particles in an regular MD simulation at 600\,K to 800\,K are about 2\,eV/\AA. In this situation, our machine learning part should try to unlearn those large contributions which is an almost insurmountable challenge. Hence, the introduction of Gaussian charges and a self-energy term bypasses this issue by normalizing the force and energy targets and greatly simplifies the training process. The machine learning short-range interaction is now solely responsible for learning all the subtle non-classical details in the PES without having to compensate large systematic errors from the long-range energy.

	Finally, a third term is optionally added for the interaction with a homogeneous external electric field $\vect{\mathcal{E}}_\text{ext}$:
	\begin{equation}
	    E_\text{ext} = - \sum_i Z_i \vect{\mathcal{E}}_\text{ext} \cdot \vect{R}_i + 2 \sum_i \vect{\mathcal{E}}_\text{ext} \cdot \vect{r}_i \label{eq:external_energy}
	\end{equation}
	or in the case of a more general external potential $V_\text{ext}$:
	\begin{equation}
			E_\text{ext} = \sum_i Z_i V_\text{ext}(\vect{R}_i) - 2 \sum_iV_\text{ext}(\vect{r}_i).
	\end{equation}

    Only classical electrostatics are explicitly present in the long-range interactions in the eMLP. In principle, dispersion interactions may still be described in such a framework, through the response of the electronic degrees of freedom to the presence of other molecules. \cite{politzer2015} However, dispersion interactions are not yet the focus of this work and we will only employ the long-range term for modeling classical electrostatic interactions effectively.

	\subsubsection{Short-range interaction} \label{sec:short-range}

    The short-range interactions are modeled using a Message Passing Neural Network (MPNN) \cite{gilmer2017neural}. More specifically, we make use of the SchNet architecture\cite{doi:10.1063/1.5019779}. In this section, we briefly elucidate its mathematical structure. For a more detailed introduction or schematic overview of the architecture, we refer to the original SchNet paper\cite{doi:10.1063/1.5019779}.

    In order to describe the full architecture, some reoccurring operations are introduced. A single dense neural network layer is given by
    \begin{equation}
        D^{N \times M}(\vect{h})= \vect{\textbf{W}} \cdot \vect{h} + \vect{b}
    \end{equation}
    in which $\vect{\textbf{W}} \in \mathbb{R}^{N \times M}$ and $\vect{b} \in \mathbb{R}^N$ are respectively the weight matrix and bias vector of the dense layer, acting on a general vector $\vect{h} \in \mathbb{R}^M$. Additionally, an activation function can be added, which we will denote by a tilde: $\Tilde{D}^{N \times M}(\vect{h}) = f(D^{N \times M}(\vect{h}))$ in which $f$ is the softplus activation function
	\begin{equation}
		f(\vect{h}) = \log\left(1 + \text{e}^{\vect{h}}\right) - \log(2).
	\end{equation}
	Next, the cutoff function is continuous and has a continuous derivative:
	\begin{equation}
		f_\text{cutoff}(r) = \begin{cases}
		1 & ,r \leq r_\text{cutoff} - r_\Delta \\
		\frac{1}{2} \left[1 + \cos\left(\pi \frac{r - (r_\text{cutoff} - r_\Delta) }{r_\Delta}\right)\right] & ,r_\text{cutoff} - r_\Delta < r \leq r_\text{cutoff} \\
		0 & ,r_\text{cutoff} < r
		\end{cases} \label{eq:cutoff_function}
	\end{equation}
	where the parameter $r_\Delta$ introduces a smooth transient zone.

    MPNN's encodes information about every particle $i$ as a feature vector $\vect{h}^t_i \in \mathbb{R}^F$ and transfers it to all its neighbors within the cutoff radius $r_\text{cutoff}$ in an iterative way. A look-up table or embedding $\vect{a}_{S_i}$, which is a trainable vector, serves as the starting point for the feature vector:
	\begin{equation}
	\vect{h}_i^0 = \vect{a}_{S_i}.
	\end{equation}
	Depending on the species $S$ of the particle $i$ (electron pair or atomic number of the atomic core), another initial vector is employed. This is the only place in the short-range contribution where a distinction between species is made. Next, messages $m^t_i$, based on the geometry of neighboring particles $j$ inside the cutoff radius, are generated in each iteration $t$
    \begin{equation}
        m^t_i = \frac{1}{J} \sum_{\substack{i \neq j \\ r_{ij} < r_\text{cutoff}}} M(\vect{h}^t_j, r_{ij}).
    \end{equation}
    The message function $M$ is implemented in SchNet\cite{doi:10.1063/1.5019779} by taking the element-wise product of the feature vector in the filter space $\mathbb{R}^G$, the so called filter-generating network $\text{W}_{\text{filter}}(r_{ij})$ and the cutoff function:
    \begin{equation}
        M(\vect{h}^t_j, r_{ij}) = D^{G \times F}(\vect{h}^t_j) \text{W}_{\text{filter}}(r_{ij}) f_\text{cutoff}(r_{ij})
    \end{equation}
    The constant $J$ is introduced to normalize the sum over all neighbors, speeding up the training of the neural network. Ideally, the constant should take the value of the average amount of neighbors of a particle. In this work, we pick $J=70$ which corresponds to the amount of neighbors in condensed systems. The filter-generating network is a simple two layered dense neural network,
	\begin{equation}
	    \text{W}_{\text{filter}}(r_{ij}) = \left[\Tilde{D}^{G \times G} \circ \Tilde{D}^{G \times N} \right] (\vect{\varphi}(r_{ij})),
	\end{equation}
	mapping $N$ radial basis functions,
	\begin{equation}
	\varphi_n = \exp\left(-\frac{(r_{ij} - \mu_n)^2}{2\sigma^2}\right)
	\end{equation}
	to the filter-space. The centers $\mu_n$ are uniformly spaced over the interval $[0, r_\text{cutoff}]$ and $\sigma = r_\text{cutoff} / N$. When all messages are calculated, the feature vectors are updated,
	\begin{equation}
	    \vect{h}^{t+1}_i = \vect{h}^{t}_i + \left[D^{F \times G} \circ \Tilde{D}^{G \times G} \right](\vect{m}^t_i),
	\end{equation}
	and the whole process is repeated $T$ times. Finally, after $T$ iterations, each feature vector is sent through the output network, yielding the particle energies:
	\begin{equation}
		\epsilon_i = \left[D^{1 \times \left\lfloor\frac{F}{2}\right\rfloor} \circ \Tilde{D}^{\left\lfloor\frac{F}{2}\right\rfloor \times F} \circ \Tilde{D}^{F \times F}\right](\vect{h}_i^T),
	\end{equation}
	while the total short-range contribution of the energy is the sum of all the particle energies
	\begin{equation}
		E_{\text{short-range}} = \sum_i \epsilon_i
	\end{equation}
	Note that the energy is translationally and rotationally invariant by construction as only the inter-particle distances enter the equations.

	\subsection{Electron localization} \label{sec:localization}

	The positioning of the electron pairs is extremely important to reproduce quantities like the dipole moment, polarizabilities and so on.
	In this work, centers of localized restricted Kohn-Sham orbitals will be used as reference data for the electron pair positions, because they are well-defined, they exactly reproduce molecular dipole moments and they provide an intuitive picture of chemical features.
	The canonical Kohn-Sham orbitals, found by solving the self-consistent equations (SCF) of Kohn-Sham density functional theory (KS-DFT) equations \cite{PhysRev.140.A1133}, are generally delocalized and have well-defined energy levels. However, any observable is invariant under unitary transformations of the occupied orbitals:
	\begin{equation}
		\ket{\psi_i} = \sum_{j} U_{ij} \ket{\phi_j}
	\end{equation}
	where $U$ is a unitary matrix, $\ket{\phi_j}$ the occupied canonical orbitals and $\ket{\psi_i}$ the new set of occupied orbitals. Any unitary transformation $U$ produces an equally valid set of orbitals. This freedom can be exploited to construct maximally local occupied orbitals.
	One of the most popular electron localization schemes was developed by Foster and Boys (FB)\cite{RevModPhys.32.300}. In this method, the unitary matrix $U$ is chosen such that the new orbitals have a minimal spatial extent, which is equivalent to minimizing the spread of all the occupied orbitals:
	\begin{equation}
		\mathcal{C}(U) = \sum_i \bra{\psi_i} \left[\vect{r} - \bra{\psi_i} \vect{r} \ket{\psi_i}\right]^2 \ket{\psi_i}. \label{eq:FB_cost}
	\end{equation}
	Because the orbitals are localized, corresponding centers are well-defined and easily computed as the expectation value of the position operator:
	\begin{equation}
		\vect{r_i} = \bra{\psi_i} \vect{r} \ket{\psi_i}
	\end{equation}
	These positions will be used as training data for the electron pair particles in eMLP.

	The centers of 1s core orbitals in second-row elements fall almost exactly on top of the corresponding nuclei, i.e.\ typically closer than $5\times10^{-4}$\,\AA. This validates our choice to combine the core electrons and the nuclei into atomic cores as a single particle.

	The FB localized orbitals correspond in most situations with chemical intuition as they agree with the Lewis structure of bonds and lone pairs in molecular systems. For instance, electron pairs can be classified as lone pairs or electrons participating in single, double or triple bonds between atoms. The electron pairs of Figure \ref{fig:water_fb_charges} are generated using this localization procedure. Note that the dipole moment of a molecule calculated with point charges located at the centers, exactly reproduces the ab-initio result. There might exist discrepancies between higher order multipoles however.

	The FB localized orbitals correspond to the ground state of the molecule. The electron pairs are situated in their equilibrium position, meaning that forces on the electron pairs are zero $\vect{f}_i = 0$. Training the eMLP to reproduce these vanishing forces will eventually lead to a model for which the equilibrium positions $\vect{r}_i^\text{eq}$ correspond with the FB positions.

	For periodic systems, Maximally Localized Wannier functions\cite{RevModPhys.84.1419} (MLWFs) replace the FB centers. MLWFs can be constructed from Bloch orbitals, for which observables are also invariant under unitary transformations\cite{PhysRevB.56.12847}. Again, the unitary transformation which minimizes the resulting spread of the orbitals is chosen, similar to the FB cost functions of Eq. \eqref{eq:FB_cost}. The resulting Wannier centers are used as reference locations of the electron pairs in periodic structures.

	\subsection{Datasets}

	\subsubsection{eQM7} \label{sec:dataset1}

	A new dataset, which we will call electron QM7 (eQM7), is created with the purpose of training and validating polarizable force fields on non-equilibrium configurations of small molecules together with external field perturbations. The QM7 dataset \cite{PhysRevLett.108.058301}, a subset of the more comprehensive GDB-13 database\cite{doi:10.1021/ja902302h}, serves as the source of the molecules, from which 6868 out of the 7165 molecules are utilized. The remaining 297 molecules contain sulfur, an element of the third row, being beyond the scope of this work. Hence, the only elements appearing in the dataset are hydrogen, carbon, nitrogen and oxygen.
	For each molecule, 500 perturbations are constructed, described in detail below, resulting in 3,434,000 different configurations in total.
	Properties of these configurations are computed with Kohn-Sham density functional theory (DFT), using the PBE0 functional \cite{doi:10.1063/1.478401, doi:10.1063/1.478522} and Aug-cc-pVTZ basis set\cite{doi:10.1063/1.462569, doi:10.1063/1.456153} in the ab-intio quantum chemistry program Psi4 \cite{doi:10.1021/acs.jctc.7b00174}. The FB localization is also performed using Psi4. After each ab-initio calculation, the following properties and arrays are stored: the total energy of the system, the positions $\{\vect{R}_i\}$ and atomic numbers $Z_i$ of the nuclei, the FB centers $\{\vect{r}_i\}$ (core electrons included), the forces on the nuclei, the forces on the FB centers (which are zero by construction) and the electric field vector.

	500 non-equilibrium configurations are generated by combining three different sampling techniques: normal mode sampling (NMS), torsion sampling and dimer sampling, which have been already successfully applied in literature \cite{Smith2019}. Unlike MD sampling, these techniques yield independent and uncorrelated structures.

	In normal mode sampling, the atoms are displaced along the normal modes of the molecule following the procedure by Smith \textit{et al.}\cite{Smith2017} In summary, the Hessian of each molecule is calculated, yielding the normal modes and afterwards the atoms are displaced along a few randomly selected modes at a temperature $T$ according to the Boltzmann distribution. The sampling is performed at high temperatures of 600\,K and 800\,K, at least double the target temperature of 300\,K. The elevated temperatures broaden the distribution of the training data, such that it includes structures with a high potential energy and a low probability at 300\,K. A machine learning force field trained without high-temperature data could erroneously underestimate the energy of structures with a low probability at 300\,K, simply due to the lack of examples. Especially for MD applications, this would be problematic\cite{Behler2021}: MD explores all low-energy regions, except those that are separated by a sufficiently high barrier from the starting point. The high-temperature data ensures such high barriers are consistently present between the regions of realistic and unrealistic structures. Because NMS cannot sample rotational barriers, torsional sampling was also employed. This is implemented by selecting a rotatable bond at random (if present) and rotating the fragment on one side of the rotatable bond over a random angle.

	Since it is not simply possible to displace individual centers of localized orbitals at will in a KS-DFT calculation, the electronic degrees of freedom are sampled by applying a homogeneous electric field across the molecule in a random direction, on top of the geometric distortions. In this way, a force acts on all the electron pairs in the direction opposite to the electric field, displacing them out of their zero-field equilibrium positions. The perturbation by a homogeneous field has some limitations. First of all, the same force acts on all electrons together, inducing more or less a collective motion such that not all normal modes of the electron pair PES are sampled independently. Second, a too strong electric field breaks the SCF convergence, especially for the larger molecules, such that only small displacements of the centers can be achieved. Even when convergence is reached, the response to strong fields may go beyond the capabilities of the local basis set used, resulting in poor wavefunctions with erroneous local orbital centers. An upper limit of 0.01\,au (=5.14$\times10^9$\,Vm$^{-1}$) for the magnitude of the electric field was put in place such that the every ab-initio calculation runs without any issue.

	To address the first limitation of the homogeneous fields, dimer configurations are constructed, in which a small probe molecule, CH$_4$, NH$_3$ or H$_2$O, exerts a realistic and more local perturbation on a molecule in eQM7. The probe molecule is placed at a random distance and orientation next to main molecule, such that the distance separating the two molecules is between 0.9 and 6\,\AA. In the same way as described above, both molecules are also subject to NMS sampling prior to creating the dimer and an additional homogeneous field is included to maximize the diversity of the training data. As a side-effect, the dimer samples also contain some information on inter-molecular interactions. However, our sampling is primarily designed to efficiently perturb centers of local orbitals, not for the calculation of precise interaction energies. The latter would require e.g.\ coupled-cluster calculations and energy differences of the form $E_{AB} - E_A - E_B$, to subtract out the relatively large intra-molecular energy changes, e.g.\ due to normal mode sampling.

	An overview of the types and numbers of perturbations applied to each molecule is given in Table~\ref{tab:overview_eQM7}.

    \begin{table}
		\begin{center}
			\begin{tabular}{ c | c}
				\textbf{Perturbation} &
				\textbf{Number of samples} \\
				\hline
                NMA@600K + Elec.\,Field &
                100 \\
                NMA@800K + Elec.\,Field &
                100 \\
                NMA@800K + Torsion + Elec.\,Field &
                150 \\
                NMA@800K + Dimer + Elec.\,Field &
                150 \\
			\end{tabular}
			\caption{Overview of the perturbations and number of samples per molecule in the eQM7 dataset. \label{tab:overview_eQM7}}
		\end{center}
	\end{table}

	\subsubsection{Beta glycine dataset}

	The dataset for $\beta$-glycine is constructed with	Quantum ESPRESSO\cite{Giannozzi_2009, Giannozzi_2017}. The self-consistent KS-DFT equations are solved using a Plane-Wave basisset and the PBE\cite{doi:10.1063/1.472933} functional with ultrasoft pseudopotentials\cite{PhysRevB.41.7892}. Grimme's dispersion corrections\cite{doi:10.1063/1.3382344} are included with Becke–Johnson damping (DFTD3-BJ)\cite{https://doi.org/10.1002/jcc.21759}. A $3\times3\times3$ Monkhorst-Pack grid\cite{PhysRevB.13.5188} for the k-points is used together with a kinetic energy cutoff of 85\,Rydberg. To make sure that the stresses correspond with the energies in a fixed k-point grid, a smooth penalty for the high energy Fourier components is introduced\cite{BERNASCONI1995501}. This is done in Quantum ESPRESSO with the following keywords: \textsc{ecfixed=80, qcutz=80} and \textsc{q2sigma=5}. The convergence threshold for the SCF equations is $10^{-8}$\,Rydberg. The positions of the electron pairs are calculated using Wannier90\cite{Pizzi_2020} and uniform electric fields are applied within the Modern theory of polarization\cite{SPALDIN20122} with the keyword \textsc{lelfield}.

	Two different sets of single point ab-initio calculations are performed. For the first set, a random electric field with a maximum strength of 0.01\,au is applied in a random direction but the stress tensor is not computed since the calculation of the stress tensor in conjunction with a non-zero external field is not implemented in Quantum ESPRESSO. In the other set, stresses are computed but, as a consequence of the previous point, no external field is applied. In this way, a total amount of 25,676 ab-initio calculations have been performed, 15,871 for the first set and 9,805 for the second set.

	In addition to the positions, the cell matrix should also be sampled extensively if the force field should be able to predict stresses. Therefore, we cannot simply use conventional normal mode sampling but follow a more general sampling strategy. First, we start by doing an ab-initio optimization of $\beta$-glycine. Next, an \textit{extended} Hessian is computed in the resulting energy minimum. This Hessian $\text{H}_\text{ext}$ is a square $3N + 9$ by $3N + 9$ matrix with $N = 20$ the number of atoms in $\beta$-glycine. The extra degrees of freedom are the 9 elements of the cell matrix. The extended Hessian has 6 zero frequencies, 3 translational and 3 rotational modes, which are discarded by projecting onto the other $3N + 3$ internal degrees of freedom. This newly created \textit{internal} Hessian $\text{H}_\text{int}$ defines the harmonic approximation of the PES:
	\begin{equation}
	    E = \frac{1}{2} \vect{x}^\text{T} \text{H}_\text{int} \vect{x}
	\end{equation}
	where $\vect{x}$ is a $3N + 3$ vector of the internal degrees of freedom. These degrees of freedom are sampled by making the connection with the Boltzmann-distribution:
	\begin{equation}
	    p(\vect{x}) \sim \exp \left( -\frac{E}{k_\text{b} T}\right) = \exp \left( -\frac{\vect{x}^\text{T} \text{H}_\text{int} \vect{x}}{2 k_\text{b} T}\right)
	\end{equation}
	Hence, the probability distribution of a sample $\vect{x}$ is a multivariate normal distribution with covariance matrix $\text{H}_\text{int}^{-1} k_\text{b}T$. After a random sample $\vect{x}$ of the internal degrees of freedom is taken, it is transformed back to the original space of $3N$ elements for the positions and $9$ for the cell matrix elements. Every configuration in our dataset is sampled using this general framework at $T = 600$\,K.

	\subsection{Data augmentation}

	Finding the location of the electron pairs or minimizing of Eq.\ \eqref{eq:minimization} assumes that the electronic energy landscape around the equilibrium positions is well-known. Therefore, the structures in our dataset were perturbed with homogeneuous electric fields with randomly sampled direction and magnitude. The magnitude was limited to 0.01\,au to avoid convergence problems in the SCF cycle. This upper limit corresponds to displacements of the electron pairs by about 0.02\,\AA\ compared to their zero-field positions. For a variety of applications, especially MD simulations, it was observed that those displacements are not large enough to sample the essential region of the electronic energy landscape. If the eMLP is trained to just these data, it will become ill-behaved when it tries to extrapolate outside the region of 0.02\,\AA\ electron pair displacements. Spurious minima and a rather chaotic PES would appear, leading to unreasonable results when minimizing the energy as function of the electron positions with Eq.\ \eqref{eq:minimization}.

	To overcome this problem, data augmentation is introduced as a preprocessing step when training the neural network. It is a popular technique in image classification to regularize the model and improve its performance\cite{Shorten2019}. In such applications, images are transformed in many different ways (flipping and rotating images, color transformations, \ldots) to artificially increase the dataset size. Every image maintains its true label after such a transformation. We utilize the same ideas to combat extrapolation of the electronic energy landscape outside the region of 0.02\,\AA. Essentially, by randomly displacing electron pairs further away from their equilibrium, a larger region of the electronic energy landscape is sampled and by training against these augmented data, the PES will become a well-behaving function outside the region of 0.02\,\AA\, displacements. However, there is one main difference with data augmentation for image classification: our true labels will change after transforming the input data. The true labels, i.e.\ energies and forces, of an augmented molecular system in which the electron pairs are displaced at random, are not known. Currently, no methodology is available to calculate the KS-DFT energy of a system where the centers of localized orbitals are chosen at will. The only thing that is known, due to the variational principle, is that the augmented system is higher in energy. Therefore, a heuristic estimate of the increase in energy, $\Delta E$, is made and the neural network is trained with the target energy $E = E_\text{gs} + \Delta E$ as its new label. Hence, outside the region of 0.02\,\AA\, not the true PES is learned, but an approximate one. This approximation is only intended to inform the neural network that no spurious minima with low energies should be predicted for large displacements of the electron pairs. The exact value of the energy increase is not critical, because the high-energy region will practically never be visited in molecular simulations, thanks to the optimization of the electron centers in Eq.\ \eqref{eq:minimization}.

	The data augmentation procedure starts by randomly selecting an electron pair and displacing it over a distance $\Delta \vect{R}$ which is uniformly sampled between 0.06\,\AA \ and 0.12\,\AA. The minimum displacement is large enough to not overlap with the sampled distribution in our dataset (i.e.\ the region of 0.02\,\AA) and the maximum displacement is small enough to still be of use when minimizing the energy around the minimum. Next, an additional energy of
	\begin{equation}
	    \Delta E = \frac{1}{2}k ||\Delta \vect{R}||^2
	\end{equation}
	is added to the target energy of the system. The numerical value of $k$ is set to be 2.0\,Ha/\AA$^2$\ and is estimated from the ab-initio results of displacing the single electron pair of H$_2$. Furthermore, an extra force $\Delta \vect{f}_i = -k \Delta \vect{R}$ pointing in the opposite direction of the displacement is assigned to the selected electron pair $i$. All the neighboring particles will also feel the influence of the displacement of the selected electron pair. Hence, the forces on these particles should also be modified such that two conditions are met: the total force is zero (for neutral systems under the influence of an constant electric field) and for non-periodic systems, the total torque should also be consistent with the external field and the electron pair displacement. In general, many solutions satisfy these constraints and in this work a unique weighted least-norm solution is always used. A full derivation of the expression of the augmented forces is given in Appendix \ref{sec:augmentation}.

	\subsection{The cost function} \label{sec:cost_functions}

	The parameters $\vect{a}$ of the neural network are trained by minimizing the following cost function:
	\begin{align}
	    \mathcal{C}(\vect{a}) = & \frac{\lambda_E}{B} \sum_{b=1}^B \left( \frac{E_\text{MLP}^{(b)}(\vect{a}) - E_{\text{tr}}^{(b)}}{N_b}\right)^2  + \frac{\lambda_f}{3N} \sum_{b=1}^B \sum_{i=1}^{N_b} \lVert \vect{F}_i^{(b)}(\vect{a}) - \vect{F}_{\text{tr},i}^{(b)}\rVert^2 \nonumber \\
        & + \frac{\lambda_c}{3C} \sum_{b=1}^B \sum_{j=1}^{C_b} \lVert \vect{f}_j^{(b)}(\vect{a}) -\vect{f}_{\text{tr},j}^{(b)}\rVert^2 + \frac{\lambda_{\sigma}}{9B} \sum_{b=1}^B \lVert \vect{\sigma}^{(b)}(\vect{a}) -\vect{\sigma}_{\text{tr}}^{(b)}\rVert^2 \label{eq:cost_function}
	\end{align}
	with $E_\text{MLP}^{(b)}(\vect{a})$ the predicted energy, $\vect{F}_i^{(b)}(\vect{a})$ the force on the $i$-th atomic core,  $\vect{f}_j^{(b)}(\vect{a})$ the force on the $j$-th electron pair and $\vect{\sigma}^{(b)}$ the stress tensor of the current system $b$ while $E_{\text{tr}}^{(b)}, \vect{F}_{\text{tr},i}^{(b)}$, $\vect{f}_{\text{tr},j}^{(b)}$ and $\vect{\sigma}_{\text{tr}}^{(b)}$ are their respective training targets. Each system $b$ contains $N_b$ atomic cores and $C_b$ electron pairs such that $\sum_{b=1}^B N_b = N$ and $\sum_{b=1}^B C_b = C$ with $B$ the total amount of systems in the current mini-batch. The adjustable weights $\lambda_E, \lambda_f$, $\lambda_c$ and $\lambda_{\sigma}$ determine respectively the relative importance between the energies, forces on the atomic cores, forces on the electron pairs and the stress tensor.

	The target energy $E_{\text{tr}}^{(b)} = E_{\text{tr,abs}}^{(b)} - E_{\text{tr,ref}}^{(b)}$, appearing in the cost function, is not the absolute ab-initio energy $E_{\text{tr,abs}}^{(b)}$ but the difference between that value and a strategically chosen reference energy $E_{\text{tr,ref}}^{(b)}$. In this way, all numerical training targets are normalized, improving the stability and speed of convergence while training the neural network. If only systems with the same chemical configuration are considered (the amount of each element and electron pairs stays the same), a single reference energy $E_{\text{tr,ref}}$ for all systems in the whole dataset will suffice. For instance, to train on the $\beta$-glycine dataset, the mean value of all the ab-initio energies $E_{\text{tr,abs}}^{(b)}$ is used as the reference energy. The same approach would not work for the eQM7 dataset because it comprises molecules with different chemical formula's. In this case, a reference energy must be defined for each chemical composition. The sum of isolated-atom energies is not a suitable reference: single atoms may have an uneven number of electrons, which is not supported by the current version of the eMLP. Instead, four reference hydrides are introduced: H$_2$, CH$_4$, NH$_3$ and H$_2$O, with corresponding energies $E_{\text{H}_2}$, $E_{\text{CH}_4}$, $E_{\text{NH}_3}$ and $E_{\text{H}_2\text{O}}$, which are meaningful in the eMLP framework. The energy of every neutral closed-shell molecule with $n_\text{H}$ hydrogen, $n_\text{C}$ carbon, $n_\text{N}$ nitrogen and $n_\text{O}$ oxygen atoms can then be expressed relative to the energies of the reference hydrides:
	\begin{equation}
	    E_{\text{ref}} = n_{\text{C}} E_{\text{CH}_4} + n_{\text{N}} E_{\text{NH}_3} + n_{\text{O}} E_{\text{H}_2\text{O}} + \frac{1}{2}(n_\text{H} - 2n_{\text{O}} - 3n_{\text{N}} - 4n_{\text{C}})E_{\text{H}_2} \label{eq:ref_energy}
	\end{equation}
	This linear combination of reference hydrides contains the same amount of atomic cores and electron pairs as the molecule. This formula is used in two places. First of all, the target reference energy $E_{\text{tr,ref}}^{(b)}$ is calculated in this way with the KS-DFT energies of the four reference molecules in Eq. \eqref{eq:ref_energy}. Secondly, the predictions being made by the eMLP are also adjusted: $E^{(b)}(\vect{a}) = E_{\text{abs}}^{(b)}(\vect{a}) - E_{\text{ref}}^{(b)}(\vect{a})$, where $E_{\text{abs}}^{(b)}(\vect{a})$ is the sum of the long-range and short-range contributions explained in Sections \ref{sec:long-range} and \ref{sec:short-range}, but here, the reference energy $E_{\text{ref}}^{(b)}(\vect{a})$ is dependent on the parameters of the neural network and is calculated with the actual predicted energies of the four reference hydrides in Eq. \eqref{eq:ref_energy}. Therefore, the four reference hydrides are included into the every mini-batch and their energies are calculated in each training step. This procedure results in a consistent calculation of energy differences over the whole training set.

	\section{Results and discussion} \label{sec:results}

	\subsection{Small molecules}

	\subsubsection{General}

	In total eight eMLP parameterizations for small molecules are trained and validated on the eQM7 dataset, four with and four without data augmentation. Besides the augmentation, the training is carried out in exactly the same way, except for the random initialization of the weights in SchNet. The training, validation and test set consist of respectively 90\%, 5\% and 5\% of the molecules. As described in Section \ref{sec:dataset1}, for each molecule 500 different ab-initio calculations were stored. After the split in train, validation and test set, all 500 calculations belonging to a single molecule stay grouped together in a single set. Hence, the validation and test set contain unseen molecules and a good performance on these sets indicate that the eMLP is not overfitting or extrapolating but instead generalizes well to other molecules.

	The cost function of Eq. \eqref{eq:cost_function} (without stresses) is minimized with the ADAM optimizer \cite{Adam} while the weights $\lambda_E, \lambda_f$ and $\lambda_c$ are all equal to one (in units of electronvolt and angstrom). The initial learning rate is $3 \times 10^{-4}$ and decays exponentially with a factor of 2 every 30 epochs. We never observe an increase in error on the validation set while training the network, mainly because the training set is large enough such that the risk of overfitting is negligible. Hence, the early stopping criterion is never triggered and each model is trained for 288 hours on V100-GPUs after which the decrease in error on the validation set becomes inconsiderable. All mini-batches contain 64 systems (the reference hydrides not included) and if data augmentation is applied, 10\% of all the systems in every mini-batch are augmented. The short-range SchNet network has $F=512$ features , $G=128$ filters, $N=32$ radial basis functions and $T=4$ interaction blocks or iterations. The cutoff radius $r_\text{cutoff}$ is 4\,\AA\ and the parameter $\gamma$ of the long-range electrostatic interaction is 0.3928\,/\AA. An overview of all the hyperparameters can be found in Table \ref{tab:hyperparameters}. Their values have been adjusted empirically.

	\begin{table}
		\begin{center}
			\begin{tabular}{ c | c}
				 \textbf{hyperparameter} & \textbf{value} \\
				 \hline
				 initial learning rate & $3 \times 10^{-4}$ \\
				 batch size $B$ & 64 \\
				 cost function weights $\lambda_E, \lambda_f$, $\lambda_c$ & 1/eV$^2$, 1 \AA$^2$/eV$^2$, 1 \AA$^2$/eV$^2$ \\
				 augmentation percentage (if applicable) & 10\% \\
				 \hline
				 cutoff radius $r_\text{cutoff}$ & 4\,\AA\  \\
				 cutoff transition width $r_\Delta$ & 0.5\,\AA\  \\
				 Gaussian charge inverse width $\gamma$ & 0.3928\,/\AA\\
				 \hline
				 features $F$ & 512 \\
				 filters $G$ & 128 \\
				 radial basis functions $N$ & 32 \\
				 interaction blocks $T$ & 4 \\
				 convolution normalization factor $J$ & 70 \\
				 \hline
				 augmentation percentage & 10\% \\
				 augmentation strength $k$ & 2\,Ha/\AA$^2$ \\
				 minimum augmentation displacement & 0.06\,\AA \\
				 maximum augmentation displacement  & 0.12\,\AA
			\end{tabular}
			\caption{Overview table of the hyperparameters of the eMLP. \label{tab:hyperparameters}}
		\end{center}
	\end{table}

	In the next two sections, we will first focus on the non-augmented models by looking at two different categories of errors: \textit{static} and \textit{dynamic} errors. Static errors are reported with the atomic cores and electron pairs at the same location of the ab-initio data. Training the model by performing gradient descent (or alternatives) on the cost function of Eq. \eqref{eq:minimization} directly minimizes this type of error. Dynamic errors on the other hand, are reported after the electron pairs are relaxed. Thus, Eq. \eqref{eq:minimization} is minimized and the errors on physical properties are reported with the electrons in their eMLP equilibrium positions, which can be slightly displaced from their true ab-initio positions. In section \ref{sec:MD_augmentation}, we will show that data augmentation is necessary when performing MD simulations. The advantages and minor disadvantages of augmented models compared to non-augmented models will be explored.

	\subsubsection{Static errors}

	In Table \ref{tab:static_errors} the static errors are reported on the test set as an average ($\pm$ standard deviation) over four models, each optimized starting from a different set of random initial model variables. Mean absolute errors (MAE) and median errors (50\% of the errors are lower than this value) are tabulated and can be compared to the intrinsic variability of the dataset. The intrinsic variability is a measure of the variance of the training data. It can be understood as the error being made by the best possible \textit{constant} model, a model that predicts the same value (i.e.\ the mean) irrespective of the input. The mean absolute error of that constant model is the intrinsic MAE. An accurate and well-performing model should have errors which are significantly lower than the intrinsic MAE. This is the case: the errors on the energies and forces are a factor 20-30 times smaller than the intrinsic MAE while the electron pair forces are about one order of magnitude smaller, showing that the eMLP is capable of making accurate predictions.

    A direct comparison with other machine learning force fields (MLFFs) is not possible since we are dealing with a new database and the model itself is different due to the inclusion of electron pair particles. A similar dataset however, is the ISO17 dataset\cite{10.5555/3294771.3294866}, on which several machine learning force fields were trained.\cite{doi:10.1021/acs.jctc.9b00181, doi:10.1021/acs.jctc.0c00347, doi:10.1063/1.5053562} It is similar because training is performed on energies and forces while the test set contains unseen molecular isomers (not contained in the training set). In those works, the nuclear forces in the validation set were reproduced with an MAE on the range of 1 to 2 kcal/mol/\AA. The eMLP has force errors just above 1 kcal/mol/\AA, putting it alongside state-of-the-art machine learning models for reliable force estimations.


	\begin{table}
		\begin{center}
			\begin{tabular}{
			    l
			    c
			    c
			    c
		    }
	            \hline
	            &
	            \multicolumn{1}{c}{energy} &
	            \multicolumn{1}{c}{forces} &
	            \multicolumn{1}{c}{electron pair forces}
	            \\
	            &
	            \multicolumn{1}{c}{[meV/atom]} &
	            \multicolumn{1}{c}{[meV/\AA]} &
	            \multicolumn{1}{c}{[meV/\AA]}
	            \\
                \textbf{Intrinsic variability} & & & \\
                \quad MAE &
                170.55 & 1106.8 & 385.9 \\
                \textbf{Non-augmented models} & & & \\
                \quad MAE &
                4.45 ($\pm$ 0.11) & 48.8 ($\pm$ 0.3) & 43.8 ($\pm$ 0.2) \\
                \quad median error &
                3.02 ($\pm$ 0.13) & 27.9 ($\pm$ 0.2) & 30.1 ($\pm$ 0.2) \\
                \textbf{Augmented models} & & & \\
                \quad MAE &
                4.95 ($\pm$ 0.25) & 52.1 ($\pm$ 0.3) & 50.3 ($\pm$ 0.4) \\
                \quad median error &
                3.45 ($\pm$ 0.39) & 29.8 ($\pm$ 0.2) & 34.6 ($\pm$ 0.3) \\
            \end{tabular}
            \caption{Static errors: electron pairs are located at the ab-initio positions. Mean absolute errors and median errors of augmented and non-augmented models are reported on the test set. Results are averaged over four different models and the value between parentheses is the standard deviation between those models. Note that the errors have been calculated after substracting the external energies of Eq. \eqref{eq:external_energy} such that the intrinsic MAE of the electron pair forces is not zero, even tough all the \textit{total} electron pair forces in the test set are zero. \label{tab:static_errors}}
        \end{center}
    \end{table}

	\subsubsection{Dynamic errors}

	In order to calculate dynamic errors, the minimization of Eq. \eqref{eq:minimization} has to be performed. Hence, here we make use of the Born-Oppenheimer eMLP energy $E_\text{eMLP,BO}$. The BFGS\cite{Nocedal2006} algorithm in SciPy\cite{2020SciPy-NMeth} is utilized to accomplish this task. The resulting dynamic errors are tabulated in Table \ref{tab:dynamic_errors}. The first three rows correspond to the average over four models on a representative randomly sampled subset of 5660 structures of the test set to reduce the computational time. It is immediately apparent that the MAE has increased multifold, while the median errors barely increase. This is due to the large tail of the error distribution. There will be a small amount of outliers, having errors orders of magnitude larger than the rest, dominating the MAE. Moreover, the standard deviation on the MAE has the same order of magnitude, indicating large fluctuations between different models. Both effects are closely linked with the error rate, also reported in Table \ref{tab:dynamic_errors}. The error rate is the fraction of the number of systems for which the optimization of the electron pairs positions fails. In those cases, the BFGS algorithm does not converge and yields solutions with non-zero electron pair forces. The particles of the model then also show erratic behavior: two or more particles are located at the same point or the particles are chaotically spread all across the molecule. For 2.4\% of all configurations, the non-augmented models cannot find a proper minimum.  In rare cases, even for structures where a solution for the electron pair positions is found, there might be large displacements compared to the ab-initio positions, giving rise to the outliers and the large MAEs. Nevertheless, in comparison to the static errors, the median errors remain almost unchanged and stable (small standard deviations). For the vast majority of molecules in the test set, energies and forces after the electron optimization are still predicted with the same accuracy as the static errors.

	The eMLP was not explicitly trained to dipole moments but is able to reproduce it when the electron pair equilibrium positions stay close to their ab-initio targets, since the ab-initio locations exactly reproduce the dipole moment of the molecule. In Table \ref{tab:dynamic_errors}, the MAE and median error on the norm of the dipole moment is reported for the structures in the test set. Again, the MAE suffers from exceptional outliers, making it less suitable to quantify the performance of the model. The median errors however, show that accurate predictions are possible with errors as low as 0.034\,D\ for the non-augmented models.

	The polarizability tensor characterizes the response of the molecular dipole moment to homogemeuous external electric field. In the final row of Table \ref{tab:dynamic_errors}, the MAE en median error on the components of the polarizability tensor are given, for each of the 343 molecules in the test. Only the polarizability tensors at the ab-initio equilibrium geometries are considered. For the non-augmented models, the median errors have a magnitude of about 0.30\,bohr$^3$, while the MAEs are much larger. Furthermore, there is a small error rate of 0.2\%. The small errors indicate that eMLP is not only capable of describing the ground-state configuration of the electron pairs, but also their response to an external field.

	A comparison of the performance of eMLP for dipole moments and polarizabilities with other models from literature is somewhat unfair: most state-of-the-art machine learning force fields train to these targets explicitly, whereas they are not explicitly included in our cost function. Still, the eMLP can be compared with models previously trained on the QM7b dataset\cite{doi:10.1063/5.0009106, Wilkins3401}. There, the errors are reported as a fraction of the intrinsic variability. Similarly, we can compute the ratio of the eMLP median errors with respect to the (median) intrinsic variabilities. A quick comparison shows that these ratios are similar (3-6\%) for both the dipole moments and polarizabilities. Note that the datasets have some differences: QM7b contains only one datapoint per molecule (7,122 versus 3,434,000 datapoints) and eQM7 in this work does not contain sulfur or chlorine.


	\begin{table}
		\begin{center}
			\begin{tabular}{
			    l
			    c
			    c
			    c |
			    c
		    }
	            \hline
	            &
	            \multicolumn{1}{c}{energy} &
	            \multicolumn{1}{c}{forces} &
	            \multicolumn{1}{c|}{dipole norm} &
	            \multicolumn{1}{c}{polarizability}
	            \\
	            &
	            \multicolumn{1}{c}{[meV/atom]} &
	            \multicolumn{1}{c}{[meV/\AA]} &
	            \multicolumn{1}{c|}{[Debye]} &
	            \multicolumn{1}{c}{[bohr$^3$]} \\
                \textbf{Intrinsic variability} & & & \\
                \quad MAE &
                174.93 & 1177 & 1.229 & 6.70 \\
                \textbf{Non-augmented models} & & & \\
                \quad MAE &
                40.8 ($\pm$ 21.1) & 104.0 ($\pm$ 23.2) & 0.169 ($\pm$ 0.042) & 0.72 ($\pm$ 0.44)\\
                \quad median error &
                3.15 ($\pm$ 0.13)& 30.0 ($\pm$ 0.6)& 0.035 ($\pm$ 0.001)& 0.26 ($\pm$ 0.01)\\
                \quad error rate &
                \multicolumn{3}{c|}{2.4\% ($\pm$ 0.8\%)} & 0.2\% ($\pm$ 0.4\%)\\
                \textbf{Augmented models} & & & \\
                \quad MAE &
                49.3 ($\pm$ 33.2)& 137.8 ($\pm$ 63.8)& 0.286 ($\pm$ 0.014) & 0.54 ($\pm$ 0.09)\\
                \quad median error &
                3.45 ($\pm$ 0.18) & 32.9 ($\pm$ 0.5)& 0.078 ($\pm$ 0.005)& 0.31 ($\pm$ 0.03)\\
                \quad error rate &
                \multicolumn{3}{c|}{0.4\% ($\pm$ 0.1\%)} & 0.0\% ($\pm$ 0.0\%)\\
            \end{tabular}
            \caption{Dynamic errors: electron pairs are optimized. Mean absolute errors and median errors of augmented and non-augmented models are reported on the test set for energies, forces and dipoles. The polarizability tensor is calculated at the ab-initio equilibrium positions of the 343 molecules in the test set such that it corresponds to a different error rate. Results are averaged over four different models and the value between parentheses is the standard deviation between those models. \label{tab:dynamic_errors}}
        \end{center}
    \end{table}

	\subsubsection{Stabilizing MD simulations with data augmentation} \label{sec:MD_augmentation}

	In this section, the applicability of eMLP to MD simulations is explored. We will primarily focus on the stability of the MD run by investigating the conserved quantity. This is a more challenging test compared to the static and dynamics errors of the previous subsections, because the nuclear motion explores a broader region in the coordinate space.

	For every molecule in the test set and for each eMLP parameterization (4 non-augmented and 4 augmented models), a 500\,fs\ NVT simulation at 300\,K\ is performed with at time step of 0.5\,fs, of which the initialization phase (first 10 steps) is discarded. Newton's equations are integrated in Yaff\cite{Yaff} with a Nosé-Hoover thermostat\cite{doi:10.1063/1.463940} with a chain length of 3. At every time step, the electron pair positions are optimized with the L-BFGS-B\cite{10.1145/279232.279236} algorithm. To speed up the convergence of L-BFGS-B, box constraints are imposed on the electron pair positions: their maximal displacement from the initial guess is limited to three times the largest nuclear displacement in the last MD step. Within these bounds a new local minimum is always found, except for some of the unstable MD runs discussed below. For each run, the conserved quantity divided by the number of atoms, $E_\text{cons}$, is characterized by two parameters. First, a line is fitted to $E_\text{cons}$ as function of time, whose slope, $s_\text{cons}$, represents the rate of conserved energy loss or gain per atom. Second, discontinuities in the BO potential energy surface are quantified by the maximum jump of the conserved quantity per atom between two timesteps, $\max |\Delta E_\text{cons}|$. Smaller values for both parameters correspond to a more stable MD run.

\begin{figure}
		\begin{center}
			\includegraphics[width=\textwidth]{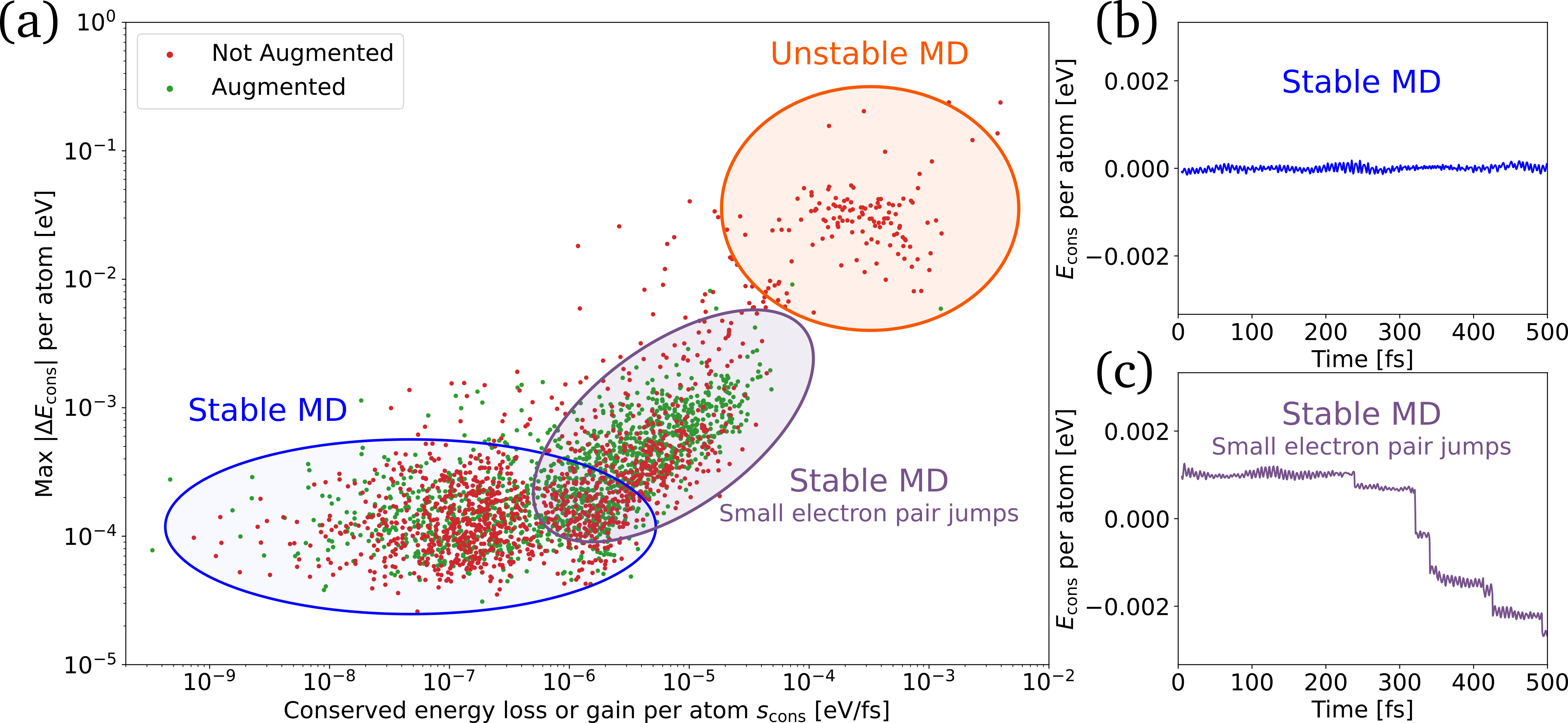}
			\caption{Panel (a): a scatter plot of the stability of MD simulations. Each point represents one MD simulation of 500\,fs\ of a single molecule, each for the 4 different augmented models (in green) and 4 non-augmented models (in red). The conserved energy per atom is tracked throughout the simulation and the value of the slope of a linear function fitted to that curve is the rate of energy loss or increase. The maximum jump in conserved energy per atom between two MD steps is given on the y-axis. Three different regions are encircled: a stable MD region, a zone with small electron pair jumps and an unstable MD region. To illustrate the behavior in the first two regions, $E_\text{cons}$ is tracked in panel (b) for stable MD and panel (c) for stable MD with small electron pair jumps.} \label{fig:scatter_md}
		\end{center}
	\end{figure}

	In Figure \ref{fig:scatter_md}(a), each MD simulation is represented by the two parameters ($s_\text{cons}, \max |\Delta E_\text{cons}|$). Every red point corresponds to a single MD run using one of the four non-augmented models. Three different regions can be distinguished in this plot: a region of stable MD simulations, a transitional region and a region with unstable MD simulations.

	Stable MD simulations are characterized by small fluctuations in the conserved quantity without a noticeable increasing or decreasing trend. Formally, these trajectories are characterized by $\Delta E_\text{cons} < 5\times10^{-4}$\,eV. A typical example is shown in Figure \ref{fig:scatter_md}(b).

	The transition region contains trajectories exhibiting sudden drops of the conserved energy per atom by 5$\times10^{-4}$ to at most 0.01\,eV, resulting in a rate of energy loss of approximately 10$^{-6}$ to 10$^{-5}$\,eV/fs. A representative example is shown in \ref{fig:scatter_md}(c). Visualization of the trajectories reveals that the electron pairs also exhibit sudden displacements when the conserved quantity drops. These jumps occur most frequently in double bonds or lone pairs, and are small enough to preserve the chemical structure of the molecule. The drop in conserved quantity corresponds the appearance of a lower local energy minimum for the electron pairs and the L-BFGS-B tends to find such lower minima whenever they are separated by only a negligible barrier from the current minimum. The resulting trajectories are still useful to sample the PES since the thermostat in MD simulations can compensate the loss or gain in conserved energy.

	Unstable MD runs are characterized by $\max |\Delta E_\text{cons}| > 0.01$\,eV, with corresponding dramatic and nonphysical rearrangements of the electronic centers. 9.8\% of the MD runs with non-augmented models show this behavior. For example, electron centers begin to overlap (which is unexpected for centers of localized orbitals) or they are ejected out of the molecule. In these cases, the eMLP wrongly assigns a lower energy to unreasonable electron configurations for which no representative training data exists.

	The results so far show that the non-augmented models may result in stable MD simulations, except when electron centers move (even briefly) outside the region sampled in the training data. The data augmentation in this work is designed to teach eMLP that it should predict a high potential energy and restoring forces (pulling the electron pairs back) whenever they venture into uncharted territory. The green points in \ref{fig:scatter_md}(a) correspond to MD simulations with the augmented models and these results confirm the effectiveness of the augmentation scheme. For all these MD runs, $\max |\Delta E_\text{cons}|$ stays below 0.01\,eV and nonphysical electron pair configurations were not observed.

	The improvement of data augmentation on MD simulations comes at a minor cost. The augmented models are trained in part to ``noisy'' labels (energies and forces) because, when the electron pairs are displaced at random, only an educated guess of the correct label can be made. For that reason, the non-augmented models have a slight edge of about 10\% for all static errors, as shown in Table \ref{tab:static_errors}. On the other hand, Table \ref{tab:dynamic_errors} compares the augmented and non-augmented models for the dynamic errors. The same behavior is again present. The median errors on energies, forces, dipoles and polarizabilities are slightly higher for the augmented models. The opposite is true for the error rates. The augmented models have an almost negligible error rate of 0.4\%, six times lower than the non-augmented models, again providing evidence that data augmentation avoids erratic solutions for the electron centers. Moreover, the error rate of the polarizabilities vanishes (only the KS-DFT ground state geometries of the molecules are considered here). In conclusion, a small amount of accuracy is traded for an increase in stability, which improves the reliability of MD simulations.

	\subsubsection{Infrared spectra}

	The computation of infrared (IR) spectra requires an excellent reproduction of both the PES of the molecule and the response of the dipole moment to changes in geometry. The static IR spectra at 0\,K consist of peaks lying at the frequencies of the normal modes. The corresponding intensities are proportional to the square of the derivative of the dipole moment along those normal modes\cite{doi:10.1021/acs.jpcc.8b08902}. The ab-initio spectra are calculated with finite differences while the spectra of the eMLP are calculated analytically by using automatic differentiation in TensorFlow\cite{tensorflow2015-whitepaper}. Only the Hessian or second order derivatives of the energy with respect to the atomic cores, electron pairs and the electric fields are necessary for the computation of the frequencies and intensities.

\begin{figure}
		\begin{center}
			\includegraphics[width=\textwidth]{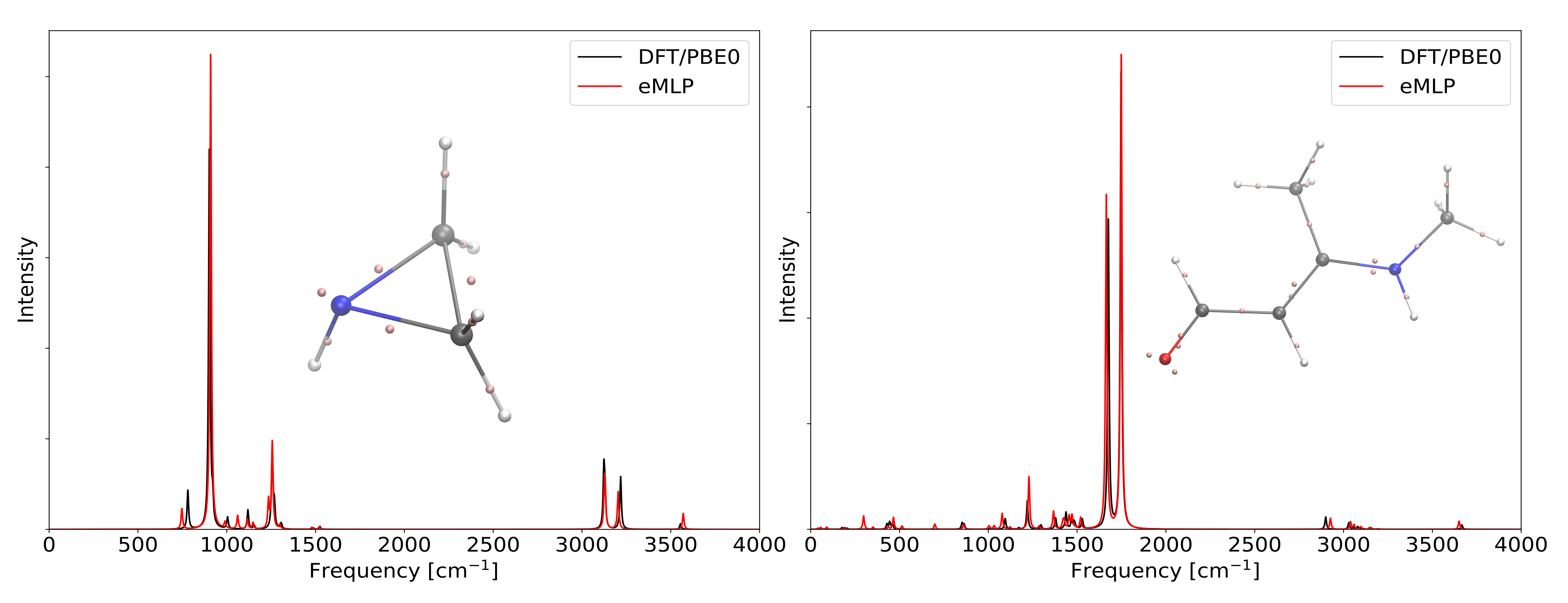}
			\caption{The IR spectra of two molecules in the test set: Aziridine (left) and 3-(Methylamino)but-2-enal (right).} \label{fig:IR_spectrum}
		\end{center}
	\end{figure}

	In Figure \ref{fig:IR_spectrum} we predict the IR spectra of two randomly selected molecules of the test set: Azaridine in the left panel and 3-(Methylamino)but-2-enal in the right panel. Similar results are obtained for all other molecules in the test set. Lorentzian line shape functions with a full width at half maximum of 10\,cm$^{-1}$ are used to visualize the spectra. Both the intensities and frequencies correspond well with the ab-initio values for all vibrational modes, including the high frequencies belonging to the C-H and N-H stretches and the low frequencies belonging to more global modes of the molecule. Furthermore the eMLP correctly identifies the peaks with the largest intensities. We emphasize that the eMLP is trained on neither molecules, showing the transferability of the eMLP for small molecules to predict IR spectra. For instance, the mean average error on all frequencies of all molecules in the validation set is approximately 15\,cm$^{-1}$, which is about the same magnitude as the full width at half maximum of the Lorentzian line shapes.

	\subsection{$\beta$-glycine}

	In this section, the eMLP will be utilized to model the response properties of $\beta$-glycine, with the main focus on the piezoelectric tensor. In essence, the piezoelectric tensor describes the coupling between mechanical properties (stress or strain) and electric properties (electric displacement or electric field). Multiple piezoelectric tensors exist, depending on the independent variables in the coupled equations. Here, we study the piezoelectric charge tensor $\vect{e}$\cite{book_AdvMechPiezo}:
	\begin{align}
	    \vect{\sigma} & = \vect{C}_{\vect{\mathcal{E}} = 0} : \vect{S} - \vect{e}^T \cdot \vect{\mathcal{E}} \\
	    \vect{D} & = \vect{e} : \vect{S} + \vect{\varepsilon}_{\vect{S} = 0} \cdot \vect{\mathcal{E}}
	\end{align}
	and the piezoelectric strain constant $\vect{d}$:
	\begin{align}
	    \vect{S} & = \vect{C}^{-1}_{\vect{\mathcal{E}} = 0} : \vect{\sigma} + \vect{d}^T \cdot \vect{\mathcal{E}} \label{eq:strain_tensor_relation}\\
	    \vect{D} & = \vect{d} : \vect{\sigma} + \vect{\varepsilon}_{\vect{\sigma} = 0} \cdot \vect{\mathcal{E}}
	\end{align}
	which both couple to the strain $\vect{S}$, stress $\vect{\sigma}$, electric field $\vect{\mathcal{E}}$ and electric displacement field $\vect{D} = \varepsilon_0 \vect{\mathcal{E}} + \vect{P}$. In these equations, $\vect{C}$ and $\vect{\varepsilon}$ are the stiffness and dielectric tensor respectively, while $\vect{P}$ is the induced dipole density. Hence,
	\begin{equation}
	    e_{ijk} = \left(\frac{\partial D_i}{\partial S_{jk}}\right)^{\vect{\mathcal{E}}} = - \left(\frac{\partial \sigma_{jk}}{\partial \mathcal{E}_i }\right)^{\vect{S}}
	\end{equation}
	\begin{equation}
	    d_{ijk} = \left(\frac{\partial D_i}{\partial \sigma_{jk}}\right)^{\vect{\mathcal{E}}} = \left(\frac{\partial S_{jk}}{\partial \mathcal{E}_i }\right)^{\vect{\sigma}}
	\end{equation}
	The the piezoelectric strain constant $\vect{d}$ cannot be calculated directly in Quantum ESPRESSO since it is impossible to compute stresses when there are electric fields being applied. Fortunately, there exists a relation\cite{book_AdvMechPiezo} between the two piezoelectric constants:
	\begin{equation}
	    \vect{d} = \vect{e} : \vect{C}^{-1}_{\vect{\mathcal{E}} = 0} \label{eq:piezo_d}
	\end{equation}
	allowing us to calculate the piezoelectric strain constant by multiplying the charge tensor with the inverse of the stiffness tensor.

\begin{figure}
		\begin{center}
			\includegraphics[width=12cm]{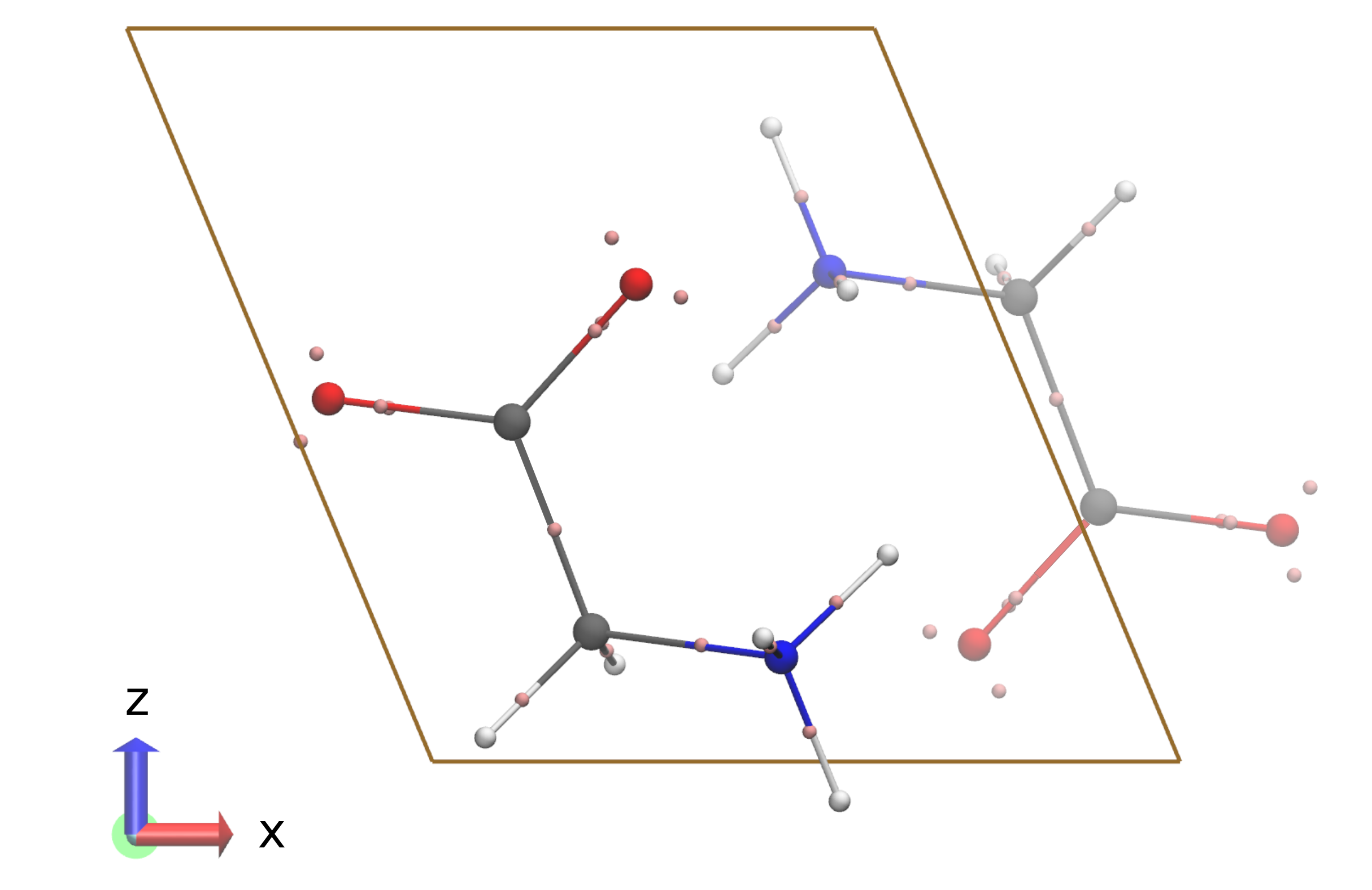}
			\caption{The ab-initio optimized structure of $\beta$-glycine. The orientation of the unit cell and the electron pairs are visualized.} \label{fig:beta_glycine}
		\end{center}
	\end{figure}

	To show that the eMLP is able to reproduce these response properties, we train a single specialized model for $\beta$-glycine. A split of 90-5-5\% is used for the training, validation and test set. Stresses are included in the cost function of Eq. \eqref{eq:cost_function} and $\lambda_E$ = 0.1/eV$^2$, $\lambda_{\sigma}$ = 0.1/GPa$^2$. The initial learning rate starts at the same value $3 \times 10^{-4}$ but decays exponentially with a factor of 2 every 1200 epochs. To calculate the periodic long-range interactions, Ewald-summation\cite{FRENKEL2002291} has been used. Only models with data-augmentation are considered here and the total training time is limited to 216 hours (early stopping is not triggered in this time-period). All remaining hyperparameters of the neural network architecture or the training algorithm have the same value as for the dataset of the small molecules. After training, the best performing model is selected and this model is used for all results below. The static errors on the test set are 3.1 meV/atom, 35 meV/\AA\, and 0.063 GPa for the MAEs on the energies, forces and stresses respectively. These are lower than the ones for the small molecules because training and validation are done on the same material, which is an easier task. This is reflected in the almost perfect reproduction of the lattice constants and volume of the unit cell, reported in Table \ref{tab:beta_glycine_lattice}. The unit cell itself belongs to space group $P2_1$ and is visualized in Figure \ref{fig:beta_glycine}.
	\begin{table}
	    \setlength{\tabcolsep}{16pt}
		\begin{center}
			\begin{tabular}{| c | c c c | c | c|}
	            \hline
				 & \multicolumn{3}{c|}{\textbf{Lattice lengths} [\AA]} & \textbf{Angles} [degrees] & \textbf{Volume} [\AA$^3$] \\
				 & $a$ & $b$ & $c$ & $\beta$ & V \\ \hline
				 DFT/PBE & 5.08 & 6.24 & 5.40 & 112.6 & 158.05 \\
				 eMLP & 5.08 & 6.25 & 5.40 & 112.4 & 158.41 \\ \hline
			\end{tabular}
			\caption{Ab-initio and predicted lattice constants, angles and the unit cell volume. \label{tab:beta_glycine_lattice}}
		\end{center}
	\end{table}

	The ab-initio values of the stiffness, dielectric and piezoelectric tensors are calculated by finite differences. We were unable to reproduce the value of 195\,pm/V for the $d_{16}$ coefficient, as reported in Guerin \textit{et. al.}\cite{Guerin2018}, presumably because of differences in computational details and dispersion model. This is not surprising, as the authors of the same paper, report that the stiffness and piezoelectric strain constants are extremely sensitive to the dispersion corrections. The stiffness constant increases if dispersion corrections are included and by making use of Eq. \eqref{eq:piezo_d}, it is obvious that the piezoelectric constants will decrease. Nevertheless, in this work we are mainly interested in the ability of the eMLP to reproduce our training data. The predicted response properties are calculated analytically by first constructing the Hessian, after which the equations of Appendix \ref{app:response_properties} are utilized.

	The diagonal components of the (ion-clamped) dielectric tensor under constant strain are reported in Table \ref{tab:beta_glycine_response_properties}. The relative error on the predicted results is 1\% or less, which shows that the response of the electron pairs to an external field is almost perfectly replicated. This is a promising result, demonstrating that the eMLP should be able to describe various dielectrics. For the remaining response properties, we make use of the Voigt notation: $xx=1$, $yy=2$, $zz=3$, $yz=4$, $xz=5$ and $xy=6$. Note that in this notation, the offdiagonal components of the strain tensor are multiplied by two: $\vect{S} = (s_1, s_2, s_3, s_4, s_5, s_6) = (s_{xx}, s_{yy}, s_{zz}, 2s_{yz}, 2s_{xz}, 2s_{xy})$. The offdiagonal components of the piezoelectric strain tensor $\vect{d}$ should also be multiplied by two since they should be counted double in the contraction of Eq. \eqref{eq:strain_tensor_relation} in Voigt notation. Table \ref{tab:beta_glycine_response_properties} shows that the diagonal components of the stiffness constants are accurately reproduced, except for the $C_{22}$ coefficient for which the deviation from the ab-initio result is slightly higher. This corresponds to direction in which the zwitterions are stacked on top of each other in Figure \ref{fig:beta_glycine}. In this direction, $\pi-\pi$ stacking between parallel molecules is expected to have a significant dispersion component, which is not yet explicitly included in eMLP. Furthermore, note that our cost function only tries to minimize the absolute (squared) errors and not the relative errors. For this reason, a minor absolute error of 1.9\,GPa on the small $C_{66}$ coefficient (which is comparable to errors of the other components), results in a more substantial relative error. In the same Table, the ab-initio values of the piezoelectric charge tensor are compared to the ones predicted by the eMLP. Due to the symmetry of the $\beta$-glycine crystal, only 8 values of the piezoelectric tensor are nonzero. All the ab-initio values, other than the $e_{22}$ coefficient, are accurately predicted. Finally, the piezoelectric strain constants are also given in Table \ref{tab:beta_glycine_response_properties}. The deviations from the ab-initio values can be attributed to the errors on the stiffness constants. For instance, the $d_{16}$ is dependent on the inverse of the $C_{66}$ coefficient, for which there is a moderate relative error, such that non negligible relative errors on the piezoelectric strain constants are unavoidable.

	\begin{table}
		\begin{center}
			\begin{tabular}{
			    l
			    S[table-format=4.3]
			    S[table-format=4.3]
			    S[table-format=4.3]
		    }
	            \hline
	            &
	            \multicolumn{1}{c}{PBE} &
	            \multicolumn{1}{c}{eMLP} &
	            \multicolumn{1}{c}{Deviation}
	            \\
                \multicolumn{4}{l}{\textbf{Dielectric constants}} \\
                \quad $\varepsilon_{11}$ &
                2.65 &
                2.68 &
                0.03\\
                \quad $\varepsilon_{22}$ &
                2.19 &
                2.21 &
                0.02\\
                \quad $\varepsilon_{33}$ &
                2.52 &
                2.55 &
                0.03\\
                \multicolumn{4}{l}{\textbf{Stiffness constants [GPa]}} \\
                \quad $C_{11}$ &
                62.4 &
                65.0 &
                2.6\\
                \quad $C_{22}$ &
                22.2 &
                31.4 &
                9.2 \\
                \quad $C_{33}$ &
                77.6 &
                84.1 &
                6.5\\
                \quad $C_{44}$ &
                8.5 &
                9.2 &
                0.7\\
                \quad $C_{55}$ &
                17.5 &
                18.9 &
                1.4 \\
                \quad $C_{66}$ &
                3.8 &
                5.7 &
                1.9\\
                \multicolumn{4}{l}{\textbf{Piezoelectric charge constant [C/m$^2$]}} \\
                \quad $e_{14}$ &
                -0.11 &
                -0.11 &
                0.00\\
                \quad $e_{16}$ &
                0.24 &
                0.24 &
                0.00 \\
                \quad $e_{21}$ &
                0.04 &
                0.04 &
                0.00\\
                \quad $e_{22}$ &
                -0.16 &
                -0.03 &
                0.13\\
                \quad $e_{23}$ &
                0.12 &
                0.10 &
                0.02 \\
                \quad $e_{25}$ &
                -0.08 &
                -0.05 &
                0.03\\
                \quad $e_{34}$ &
                0.01 &
                0.02 &
                0.01 \\
                \quad $e_{36}$ &
                -0.05 &
                -0.05 &
                0.00\\
                \multicolumn{4}{l}{\textbf{Piezoelectric strain constant [pm/V]}} \\
                \quad $d_{14}$ &
                -25.0 &
                -19.8 &
                5.2\\
                \quad $d_{16}$ &
                71.6 &
                46.5 &
                25.1 \\
                \quad $d_{21}$ &
                1.0 &
                0.3 &
                0.7\\
                \quad $d_{22}$ &
                -9.6 &
                -1.8 &
                7.8\\
                \quad $d_{23}$ &
                1.9 &
                1.2 &
                0.7 \\
                \quad $d_{25}$ &
                -4.3 &
                -1.3 &
                3.0\\
                \quad $d_{34}$ &
                3.2 &
                3.9 &
                0.7 \\
                \quad $d_{36}$ &
                -15.1 &
                -9.9 &
                5.2\\
                \hline
			\end{tabular}
			\caption{The ab-initio PBE and eMLP prediction of the dielectric, stiffness and piezoelectric constants in $\beta$-glycine and the deviation between those two values. \label{tab:beta_glycine_response_properties}}
		\end{center}
	\end{table}

	\section{Conclusion and outlook} \label{sec:conclusion}

	In this work, we introduced the eMLP, a new explicit electron force field making use of machine learning for short-range interactions combined with classical electrostatics at longer ranges. Centers of localized Foster-Boys or Wannier orbitals serve as training data for the positions of electron pair particles in the eMLP, which has several advantages. These centers provide extensive reference data, they exactly reproduce molecular dipole moments, intuitively represent chemical features and are well defined. Two new datasets were created to showcase eMLP's capabilities and performance. The eQM7 dataset consists of a variety of independently sampled configurations for each of the 6,868 small molecules in the dataset. The $\beta$-glycine dataset uses a generalized version of normal mode sampling to sample configurations with different unit cells and fractional coordinates. It was shown that force errors under 0.05\,eV/\AA\, can be achieved, even after (re)optimizing the electron pairs with a trained eMLP. Furthermore, IR-spectra of unseen molecules are predicted accurately. For $\beta$-glycine, the eMLP is able to model elastic, dielectric and piezoelectric responses, which are hard to accomplish with conventional force fields. These test cases demonstrate the potential of eMLP for the simulation of physical properties involving non-trivial electronic behavior. To run MD simulations, it is necessary to train eMLP with data augmentation, a technique in which an electron center is displaced over a larger distance with a large associated energy increase. Such data cannot be generated with electronic structure calculations but are needed in the training set to prevent extrapolation issues.

	During the development of eMLP, new challenges arose for which future methodological advances are of interest. In this work, only relatively weak homogeneous electrical fields were used to perturb the centers of localized orbitals, which provides an incomplete picture on the electronic response function. Kohn-Sham DFT data for larger and more diverse displacements of the centers would be a valuable addition to the training set and may eventually replace the data augmentation in this work.

	Several extensions or more complex use-cases of the eMLP can be realized in future developments. Up until now, the electronic degrees of freedom were modeled as electron pairs but a subdivision in a separate class of spin-up and spin-down electrons, in analogy to LEWIS$\bullet$, will enable the simulation of radicals or magnetic systems. The explicit treatment of long-range dispersion interactions can also be beneficial. Moreover, the short-range interactions can be modeled by any state-of-the-art machine learning force field, especially when more data-efficient or accurate models become available. Also the simulation of chemical reactions with the eMLP should be investigated, since this could potentially enable a natural description of redox reactions and charge-transport phenomena.

	\section*{Data and software availibility}

	The eQM7 dataset is available on the Materials Cloud Archive (\url{https://doi.org/10.24435/materialscloud:66-9j}). The $\beta$-glycine dataset is also available on the Materials Cloud Archive (\url{https://doi.org/10.24435/materialscloud:jn-44}). A reference implementation of the eMLP is available on github at \url{https://github.com/mcoolsce/eMLP} and a release is archived on Zenodo (\url{https://doi.org/10.5281/zenodo.5526796}).

	\begin{acknowledgement}

	This work is supported by the Fund for Scientific Research Flanders (FWO, grant no. 11D0420N). The work is furthermore supported by the Research Board of Ghent University (BOF). The computational resources (Stevin Supercomputer Infrastructure) and services used in this work were provided by the VSC (Flemish Supercomputer Center), funded by Ghent University, FWO and the Flemish Government – department EWI.

	\end{acknowledgement}

	\appendix

	\section{Analytical expressions of eMLP response properties} \label{app:response_properties}

	Here, we give a brief overview of the necessary equations to calculate responses properties analytically with eMLP for solids and molecules. The evaluation of derivatives appearing in these expressions is implemented with automatic differentiation. The starting point is the full extended Hessian with respect to the fractional coordinates of both atomic cores and electron pairs, the elements of the unit cell and the electric field:
	\begin{equation}
	    \text{H}_{\text{extended}} = \begin{pmatrix}
	    \text{H}_{\text{ff}} & \text{H}_{\text{fa}} & \text{H}_{\text{fe}} \\
	    \text{H}_{\text{af}} & \text{H}_{\text{aa}} & \text{H}_{\text{ae}} \\
	    \text{H}_{\text{ef}} & \text{H}_{\text{ea}} & \text{H}_{\text{ee}}
	    \end{pmatrix}
	\end{equation}
	The subscripts denote the respective groups of derivatives of the Hessian:
	\begin{description}
	    \item[f] : Towards fractional coordinates of the $N$ atomic cores and  $C$ electron pairs (3$N$ + 3$C$ elements in total).
	    \item[a] : Towards elements of the unit cell (9 elements in total). Usually two indices are used to identify rows (for cell vectors) and columns (for components of cell vectors).
	    \item[e] : Towards the electric field (3 elements).
	\end{description}
	For instance, using this notation, $\text{H}_{\text{fa}} = \text{H}_{\text{af}}^T \in \mathbb{R}^{(3N + 3C) \times 9}$ is the off-diagonal block of the Hessian with first index the fractional coordinates and the second index the unit cell elements. In the following equations, the unit cell matrix is $A_{ij}$ where the \textit{rows} are the lattice vectors. All the following response properties are calculated at 0\,K after an optimization of the structure such that the Hessian has no negative eigenvalues and is invertible. For the stiffness tensor, we find
	\begin{equation}
	    C_{ijkl} = \frac{1}{V} \sum_{mn} A_{mi}A_{nk} \left(\text{H}_{aa} - \text{H}_{af}\text{H}_{ff}^{-1}\text{H}_{fa} \right)_{mjnl}
	\end{equation}
    where the second term is a consequence of the relaxation of the fractional coordinates, calculated using vibrational subsystem analysis (VSA)\cite{doi:10.1063/1.3013558}. Furthermore, the term in parenthesis is a $9 \times 9$ matrix of which the elements are reordered in a  $3\times 3 \times 3 \times 3$ tensor to perform the contraction over $m$ and $n$. The total polarizability of a molecule is simply
    \begin{equation}
        p_{ij} = \left( \text{H}_{ef}\text{H}_{ff}^{-1}\text{H}_{fe} \right)_{ij}.
    \end{equation}
	This formula also includes atomic core relaxations. These can be frozen as well, by only taking the sub matrices corresponding to the electron pairs into account. The latter is reported in the main text. The (relative) dielectric constant is
    \begin{equation}
        \varepsilon_{ij} = \delta_{ij} + \frac{1}{V \varepsilon_0} \left( \text{H}_{ef}\text{H}_{ff}^{-1}\text{H}_{fe} \right)_{ij}
    \end{equation}
    with $V$ the volume of the unit cell. Again, by excluding the atomic cores in the sum matrices, one can compute the so-called ion-clamped static dielectric tensor, which is reported in the main text. Next, the piezoelectric charge constant is
    \begin{equation}
        \widetilde{e_{ijk}}= \frac{1}{V} \sum_{m} A_{mj}\left( \text{H}_{ef}\text{H}_{ff}^{-1}\text{H}_{fa} \right)_{imk}
    \end{equation}
    where again the matrices are reshaped at the end. Note that this is the \textit{proper} piezoelectric tensor $\widetilde{e_{ijk}}$, which can be measured experimentally\cite{VANDERBILT2000147}. Furthermore, the proper piezoelectric tensor has the correct symmetry and is invariant under equivalent displacements of the particles. Indeed, displacing a particle over an integer multiple of the lattice vectors (putting it in a neighbouring cell), should not affect any observable quantities. The dipole vector itself is not invariant under such a displacement, which is not an issue as it cannot be measured experimentally for periodic systems. Only the changes of the dipole vector, due to internal relaxations of the particles, can be measured experimentally. The proper and improper piezoelectric tensor $e_{ijk}$ are related by
	\cite{VANDERBILT2000147}:
	\begin{equation}
	\widetilde{e_{ijk}} = e_{ijk} + \delta_{jk} P_i - \delta_{ik}P_j
	\end{equation}
    where $P_i$ is the dipole density. The improper piezoelectric tensor $e_{ijk}$ is not symmetric in general and depends on the unit cell under consideration. Only the proper piezoelectric tensors are reported in the main text.

	\section{Derivation augmented forces} \label{sec:augmentation}

	In the augmentation procedure, the electron pair $i$ is displaced over a distance $\Delta \vect{R}_i$ and as a result, it receives an additional force $\Delta\vect{F}_i$, which wants to push the particle back to its original position:
	\begin{equation}
	    \Delta\vect{F}_i = -k \Delta \vect{R}_i
	\end{equation}
	In the following discussion, the index $j$ is used for all particles, atomic cores and electron pairs. Since the only external force on the system is a constant electric field $\vect{\mathcal{E}}$, the net force on a electrically neutral system must be zero:
	\begin{equation}
	    \sum_j \vect{f}_{j'} = 0 \label{app:force_balance}
	\end{equation}
	and the total torque must be equal to:
	\begin{equation}
	    \sum_j \vect{r}_{j'} \times \vect{f}_{j'} = \vect{d}' \times \vect{\mathcal{E}} \label{app:torque_balance}
	\end{equation}
	with $\vect{d}'$ the dipole vector of the system, $\vect{f}_{j'}$ and $\vect{r}_{j'}$ the forces and positions of the particles \emph{after} the electron pair has been moved. The goal is to find the extra forces $\Delta \vect{f}_j$ for $j \neq i$ such that equations \eqref{app:force_balance} and \eqref{app:torque_balance} are fulfilled. This results in the following equations for $\Delta \vect{f}_j$:
	\begin{equation}
	\begin{cases}
	    \sum_{j \neq i} \Delta \vect{f}_j = -\Delta\vect{F}_i \\
	    \sum_{j \neq i} \vect{r}_j \times \Delta \vect{f}_j = q_i \Delta \vect{R}_i \times \vect{\mathcal{E}} - \Delta \vect{R}_i \times (\vect{f}_i + \Delta\vect{F}_i) - \vect{r}_i \times \Delta\vect{F}_i \label{app:balance}
	\end{cases}
	\end{equation}
	Note that the equations are invariant to a global translation. This property will be used later on. If one also demands that the extra forces are as small as possible due to the disruption caused by the augmentation, one should find the stationary point of the following Lagrangian:
	\begin{equation}
	    \mathcal{C}(\{\Delta \vect{f}_j\}) = \frac{1}{2} \sum_{j \neq i} \phi(r_{ji}) ||\Delta \vect{f}_j||^2  + \vect{\lambda} \cdot \left(-\Delta\vect{F}_i -  \sum_{j \neq i} \Delta \vect{f}_j\right) + \vect{\mu} \cdot \left(\vect{M} - \sum_{j \neq i} \vect{r}_j \times \Delta \vect{f}_j\right)
	\end{equation}
	where $\vect{\lambda}$ and $\vect{\mu}$ are two vectorial Langrange multipliers. The vector $\vect{M} = q_i \Delta \vect{R}_i \times \vect{\mathcal{E}} - \Delta \vect{R}_i \times (\vect{f_i} + \Delta\vect{F}_i) - \vect{r}_i \times \Delta\vect{F}_i$ is introduced to simplify the notational burden and $\phi(r_{ji})$ is a possible weight factor depending on the distance between the particles $i$ and $j$. Minimization with respect to $\Delta \vect{f}_j$ yields
	\begin{equation}
	   \phi(r_{ji}) \vect{f}_j - \vect{\lambda} - \vect{\mu} \times \vect{r}_j = 0 \\
	\end{equation}
	or
	\begin{equation}
	    \vect{f}_j = \frac{\vect{\lambda} + \vect{\mu} \times \vect{r}_j}{\phi(r_{ji})} \label{app:augmented_force}
	\end{equation}
	Maximizing with respect to the langrangian multipliers and substituting this result, yields
	\begin{equation}
	    \begin{cases}
	        \vect{\lambda} Q_0 + \vect{\mu} \times \vect{Q}_1 = -\Delta\vect{F}_i \\
	        \vect{Q}_1 \times \vect{\lambda} + \overline{\overline{\text{Q}}}_2 \vect{\mu} = \vect{M}
	    \end{cases}
	\end{equation}
	with $Q_0 = \sum_{j \neq i} \frac{1}{\phi(r_{ji})}$, $\vect{Q}_1 = \sum_{j \neq i} \frac{\vect{r}_j}{\phi(r_{ji})}$ and the 3-by-3 tensor $\overline{\overline{\text{Q}}}_2 = \sum_{j \neq i} \frac{||\vect{r}_j||^2 \text{I} - \vect{r}_j\vect{r}_j}{\phi(r_{ji})}$. Next, consider following translation: $\vect{r}_j \rightarrow \vect{r}_j + \vect{t}$. For the following choise of the translation vector,
	\begin{equation}
	    \vect{t} = -\frac{\vect{Q}_1}{Q_0},
	\end{equation}
	$\vect{Q}_1$ in the new coordinate system will be zero, simplifying the equations. Furthermore, the vector $\vect{M}$ should now also be calculated in this coordinate system, which is valid since Eq. \eqref{app:balance} has the same form after a translation. Hence,
	\begin{equation}
	    \begin{cases}
	    \vect{\lambda} = -\frac{\Delta\vect{F}_i}{Q_0} \\
	    \vect{\mu} = \overline{\overline{\text{Q}}}_2^{-1} \vect{M}
	    \end{cases}
	\end{equation}
	Finally, substituting these in Eq. \eqref{app:augmented_force} gives us the sought for answer. In this work, the following weight function is being used, yielding larger forces for particles closer to the displaced electron pair:
	\begin{equation}
	    \phi(r_{ji}) = \frac{r_{ji}^2}{f_\text{cutoff}(r_{ji})}
	\end{equation}
	with $f_\text{cutoff}(r_{ji})$ the cutoff function of Eq. \eqref{eq:cutoff_function}. (The inverse of $\phi$ appears in the solution, such that forces on particles outside the cutoff sphere of electron pair $i$ become zero.)


	\bibliography{bibliography}

\end{document}